\documentclass[aps.pra,twocolumn,showpacs,preprintnumbers,amsmath,amssymb]{revtex4}
\usepackage[utf8]{inputenc}
\usepackage[english]{babel}

\usepackage{amsmath}
\usepackage{amssymb}

\usepackage{graphicx}
\usepackage{placeins}

\graphicspath{{./}}

\usepackage[figurename=FIG., justification=RaggedRight]{caption}
\usepackage{subcaption}
\usepackage{physics}
\usepackage[titletoc]{appendix}

\usepackage{amsmath}
\usepackage{amssymb}
\usepackage[]{qcircuit}

\begin{document}

\title{Hybrid digital-analog simulation of many-body dynamics with superconducting qubits}
\author{D. V. Babukhin$^{1,2}$, A. A. Zhukov$^{1,3}$, W. V. Pogosov$^{1,4}$}

\affiliation{$^1$Dukhov Research Institute of Automatics (VNIIA), 127055 Moscow, Russia}
\affiliation{$^2$Russian Quantum Center (RQC), 143026 Moscow, Russia}
\affiliation{$^3$National Research Nuclear University (MEPhI), 115409 Moscow, Russia}
\affiliation{$^4$Institute for Theoretical and Applied Electrodynamics, Russian Academy of
Sciences, 125412 Moscow, Russia}

\begin{abstract}
 In recent years, there has been a significant progress in the development of digital quantum processors. The state-of-the-art quantum devices are imperfect, and fully-algorithmic fault-tolerant quantum computing is a matter of future. Until technology develops to the state with practical error correction, computational approaches other than the standard digital one can be used to avoid execution of the most noisy quantum operations. We demonstrate how a hybrid digital-analog approach allows simulating dynamics of a transverse-field Ising model without standard two-qubit gates, which are currently one of the most problematic building blocks of quantum circuits. We use qubit-qubit crosstalks (couplings) of IBM superconducting quantum processors to simulate Trotterized dynamics of spin clusters and then we compare the obtained results with the results of conventional digital computation based on two-qubit gates from the universal set. The comparison shows that digital-analog approach significantly outperforms standard digital approach for this simulation problem, despite of the fact that crosstalks in IBM quantum processors are small. We argue that the efficiency of digital-analog quantum computing can be improved with the help of more specialized processors, so that they can be used to efficiently implement other quantum algorithms. This indicates the prospect of a digital-to-analog strategy for near-term noisy intermediate-scale quantum computers.
\end{abstract}

\maketitle
\date{\today }

\section{Introduction}

Quantum computing is a paradigm of solving computation problems using a quantum physical system as a processor. This paradigm initiated from the idea to use one quantum system to simulate another quantum system, proposed at the end of the twentieth century in works of Manin \cite{Manin} and Feynman \cite{Feynmann}. A decade later Lloyd \cite{Lloyd1996} argued that it is possible to simulate the dynamics of every local quantum system in a quantum-computational way.

In parallel with this, there were attempts to apply the quantum computing paradigm to non-physical computation tasks \cite{Deutsch1992}, which led to the first potentially-useful quantum algorithms with a speed-up (factorization of a number \cite{Shor1994} and search in an unstructured database \cite{Grover1996}). Afterward, the development of quantum computation studies progressed rapidly, and more algorithms have been proposed, e.g., solving a system of linear equations \cite{HHL} or a quantum principal component analysis \cite{QPCA}.

Searching for algorithms with quantum speed-up led to the notion of "quantum supremacy" - an ability of quantum computers to solve specific problems with significantly fewer resource requirements compared to the classical computers \cite{Supremacy}. Recently an experimental realization of quantum supremacy \cite{Arute2019} heralded a new stage of quantum computing technology.

Another achievement of the last decade is the transition of quantum computer prototypes from experimental laboratories to the public. Several examples of quantum processors are available for the development of quantum-enhanced applications and research purposes for users all over the world \cite{ibmqx,rigetti,ionq}.
These are noisy intermediate-scale quantum (NISQ) devices \cite{NISQ}, characterized by a relatively small number of qubits, limited qubit-qubit connectivity, faulty single- and two-qubit gates and qubit state readout procedures, as well as low coherency times and the presence of crosstalks between qubits, leading to uncontrollable dynamics during the computation process.
In the present state, quantum computers are far away from fault-tolerant requirements \cite{QEC_beginners}, and there is much to be done in this emerging field. Nonetheless, there are first practically-interesting results, many scientifically-interesting works, as well as attempts to compensate the imperfection of technology with additional techniques, see, e.g., Refs. \cite{variat,nucleus,Behera2017,Huffman2017, Ku2019, sotnikov2019neural, Gangopadhyay2018,Doronin2020,mitig1,mitig2,mitig3,mitig4}, which inspire hope in the successful future of quantum computing.

One of the promising applications of quantum computing is the calculation of various characteristics of quantum many-body systems. Wave functions of these systems reside in an exponential-size Hilbert space. Due to exploding dimensionality, it is practically impossible for classical computers to calculate properties of such systems from first principles even for a medium particle number. With quantum computing, it becomes realistic to operate with exponentially-many quantum states in a controllable way. Having a properly defined observable, a significant speed-up of computation is possible.

There exist several approaches for modeling properties of quantum mechanical systems using a quantum device. The digital approach implies preparing an initial quantum state and constructing evolution operator using of single- and two-qubit gates from a universal set or more specialized gates preferable for a particular class of problems that can enhance compilation efficiency \cite{foxen2020demonstrating}. This approach requires essentially a universal quantum computer, which has qubits' and operations' quality high enough to guarantee a desired precision. This is the most involved kind of quantum computation, but it can be applied to the widest range of solvable problems \cite{Ladd2010, Georgesce2010}. An alternative method within digital strategy is associated with variational techniques. The analog approach implies constructing a quantum mechanical system with a sufficient level of control, whose dynamics can mimic the dynamics of other quantum systems. The field of analog quantum simulators was developing alongside with universal quantum computing in the last decades, being an interesting alternative  \cite{Buluta2009, Georgescu2014}.
The idea to use a system of qubits, connected with tunable couplers, to solve optimization problems with quantum annealing was studied extensively. Although being suitable for narrow range of problems, this approach may be one of the first to become industrially applicable example of quantum computation \cite{Harris2010, Weber2017}.

There is a hybrid digital-analog approach to quantum computation, which until now has attracted much less attention, but potentially can incorporate the best of two worlds, see, e.g., theoretical studies \cite{SB,Hu2007,DAQC1}.
In this approach, a register of qubits is controlled by single-qubit operations from one hand, and by evolution under a Hamiltonian, which provides quantum entanglement, from the other hand. The interaction Hamiltonian is embedded in quantum hardware on a physical level due to the couplings between qubits. The interaction can be tunable, i.e., it can be switched off during execution of single-qubit gates by using dynamically adjustable couplers or frequency-tunable qubits. It is, however, technically simpler to use always-on interaction and fast single-qubit gates, whose fidelity is thus only slightly decreased by the interaction. In principle, qubit-qubit interaction together with single-qubit gates allow for the realization of an arbitrary unitary transformation \cite{SB} including basic subroutines of quantum computation such as quantum Fourier transform \cite{DAQC2}, but it is more straightforward to use it to specialized problems such as quantum simulation. One of the ideas is to create initial highly excited state of qubits using single-qubit gates and then to let the system to evolve under the interaction. This idea was realized experimentally for superconducting qubits \cite{Simulation1} as well as for trapped-ion systems \cite{Simulation4,Simulation2}. The hybrid quantum computational strategy, which relies on single-qubit control and interaction between qubits, is a perfect candidate to use in applications of near-term NISQ devices, since one of the most problematic features of digital quantum hardware, which is a need for high-fidelity two-qubit operations of the universal set, is not required within this strategy. For specific (although wide) class of problems, using digital-analog approach can lead to faster transfer of quantum computation applications into the real life.

In this paper, we show an example of the digital-analog simulation by using a rather unexpected feature of superconducting quantum processors with fixed frequency qubits and illustrate our ideas with superconducting quantum processors of IBM Quantum Experience. We use residual stray couplings between the qubits connected by resonators, which are known as crosstalks, to simulate the dynamics of spin clusters. Let us emphasize that the crosstalks are responsible for additional errors in digital quantum computation. In contrast, we here use them to simulate many-body systems, described by Hamiltonians with interaction terms similar to these couplings.
Within this approach, we avoid the execution of two-qubit quantum gates, thus mitigating error accumulation problem. The role of crosstalks in our work is to provide always-on interaction between qubits. In contrast to Refs. \cite{Simulation1,Simulation4,Simulation2}, we rely more deeply on a digital side of the hybrid strategy and consider Trotterized (discretized) dynamics instead of a fully analog evolution. In essence, we apply single-qubit gates periodically in time and let the system to evolve freely between them. Trotterized quantum simulation potentially yields much better flexibility in terms of the Hamiltonian we wish to simulate.

We observe a significant improvement of the computation accuracy in simulating dynamics of spin clusters, when the digital-analog approach is used instead of fully digital approach based on controlled-NOT (CNOT) gates. Notice that in these studies we are mostly interested
in infidelities between perfect and imperfect realizations of the same
circuit within the digital and digital-analog approaches, i.e., in both cases we use fixed number of Trotter steps, which is dictated by the maximum evolution time considered. We also address quantum Fourier transform and compare accuracies these two approaches which turn out to be nearly the same for this particular quantum circuit. We stress that these results were obtained using available quantum processors with crosstalks being intentionally suppressed. A quantum hardware can be optimized to improve the quality of such a hybrid quantum computation, based on always-on interaction (keeping individual addressability of qubits). We provide simple estimates for the optimal interaction constant between qubits and show that crosstalks in IBM Q machines are one or two orders of magnitude weaker; therefore the efficiency of quantum computation can be further greatly improved (at least, for certain quantum algorithms). This implies that specialized quantum processors based on digital-analog strategy and always-on interaction are prospective within NISQ era.

In digital quantum computers always-on interaction produces non-markovian dynamics of qubits, which gives rise to correlated errors. The character of non-markovian dynamics in IBM Q quantum machines and its impact on digital computation are also discussed in the present paper.

This paper is organized as follows: in Sec. II, we provide background about origin of crosstalks between superconducting qubits as well as describe our tools for modeling the evolution of many-body systems. In Sec. III, we describe experimental results for the dynamics of spin clusters. Sec. IV deals with summary and discussion.

 \section{Background and Context}
 \subsection{Superconducting fixed-frequency qubits and the origin of their residual interaction}

Nowadays, superconductivity-based quantum processors are the most widespread devices for quantum computing. The superconducting qubits can have both fixed and tunable frequencies. All existing realizations are imperfect and gathered under the name noisy intermediate-scale quantum devices. In order to implement digital-analog quantum computation ideas, it is necessary to have an interaction between qubits together with single-qubit gates and readouts. The interaction can be provided either by capacitive coupling between the qubits or couplings via microwave resonators and it can be either static or tunable. In this paper, we focus on digital-analog computation with IBM quantum devices and then we briefly discuss optimal regimes of coupling strengths for the realization of the hybrid strategy.

IBM Q quantum processors operate with fixed-frequency qubits and coupling between them is due to microwave resonators. These processors are built on transmons - superconducting qubits, which consists of a Josephson junction shunted by a large capacitor \cite{GambettaSC}. The resonance frequency of these devices is typically 4-6 GHz. In terms of a lumped circuit model, it is a parallel LC oscillator with the nonlinear inductor and a linear capacitor. The qubit is two levels of the resulting anharmonic oscillator. For selective control of transitions between energy levels and effective separation of the qubit Hilbert space, anharmonicity of this oscillator is made sufficiently large. Qubit frequencies are fixed to have long coherence times, which approach 50-100 $\mu s$. Full single-qubit control is carried out by applying external microwave driving through a transmission line strongly coupled to qubits. Every qubit is connected to the output resonator, which allows doing measurements of the qubit state. There are qubits connected with a resonator in dispersive regime (i.e., when a frequency of the resonator is significantly detuned from qubits frequencies), between which a two-qubit gate can be applied via microwave pulses.

In general, qubits coupled with a resonator interact with virtual photons (even in the absence of controlling pulses), and this interaction can be described by the Dicke model \cite{GambettaSC, EffectiveH}.
Since the frequency of the photon is significantly larger than the excitation frequencies of qubits, photons can be eliminated from the description of the system using perturbation theory that gives rise to the effective interaction between qubits \cite{GambettaSC, EffectiveH}. Thus, the degrees of freedom, related to physical qubits, slightly hybridize with each other.

To keep an individual addressability of qubits given the hybridization, a modified set of two-level systems can be introduced, which leads to the diagonalization of the Hamiltonian in the new basis \cite{GambettaSC}.
These new qubits, which form computational basis, are constructed from transmons degrees of freedom with small contributions from the other transmons connected with the given one through resonators.
However, transmons cannot be treated as ideal two-level systems due to limited anharmonicity. Higher levels of transmons can be populated via transitions from the first excited state. The incorporation of higher levels into the effective description results in the additional interaction of Ising $ZZ$ type in the Hamiltonian of qubits in the computational basis \cite{GambettaSC}. The interaction between two qubits 1 and 2 in this representation is given by the term of the form $-J^{phys}_{1,2} \sigma_1^z \sigma_2^z$, where  $J^{phys}_{1,2} \sim (\delta_1+\delta_2)$, $\delta_{1,2}$ being anharmonicities of qubits 1 and 2; $J^{phys}_{1,2}$ also depends on detunings between various frequencies \cite{EffectiveH}.

For a quantum processor build from fixed-frequency qubits, a connectivity map of qubit-qubit residual interaction coincides with the connectivity map of the two-qubit gates of the processor. Let us stress that this interaction is static and exists even when gates are not active. The values of $J_{ij}^{phys}$ between different qubits are provided for all quantum devices of the IBM Quantum Experience project \cite{ibmqx1} and typical $J_{ij}^{phys}$ is 50-100 kHz, all of them being positive.
Values of crosstalks are measured with a modification of Bilinear Rotational Decoupling, which is a special technique known from NMR experiments (see the device documentation for IBM processors \cite{ibm_device}). It is based on the Ramsey experiment with echo pulses, applied to a pair of qubits, where one qubit state is measured at the end. This experiment is performed for two initial states of the qubit, and the final rotation angle is varied to observe a difference in oscillations for two experiments, which is induced by the crosstalk \cite{ibm_device}.
It turns out that $1/J_{ij}^{phys}$ is usually several times smaller than both $T_1$ and $T_2$, which implies that it is possible to perform few entangling operations using embedded $ZZ$ interaction before the decoherence would play a significant role. For example, for 5-qubit chip IBM QX2 $\overline{J_{ij}^{phys}} \overline{T_1} \approx 4.3$ and $\overline{J_{ij}^{phys}} \overline{T_2} \approx 3.8$, where averaging is performed over all qubits of the device. For 14-qubit chip IBM QX14 $\overline{J_{ij}^{phys}} \overline{T_1} \approx 2.6$ and $\overline{J_{ij}^{phys}} \overline{T_2} \approx 3.6$.

All these estimates are made for the available quantum hardware, which is optimized to keep crosstalks small. We believe that it is possible with a specialized hardware to improve these estimates, i.e., to increase typical values of $J_{ij}^{phys}T_{1,2}$ and to enable for the larger numbers of entangling steps of the algorithm. Of course, this also implies a trade-off between the individual addressability of qubits and the increase of crosstalks. This issue is discussed in a more detail in Section IV.

Strictly speaking, effective Hamiltonian for fixed-frequency qubits in IBM Q architecture contains not only additional terms of ZZ type but also terms of other sorts (such as ZX), which are, however, subdominant within a perturbative expansion \cite{EffectiveH}. If we disregard them in our computation scheme, they produce certain errors. In principle, such errors can be suppressed digitally, i.e., by appropriately applying single-qubit gates and modifying an algorithm itself. However, we are not going to pursue this issue in the present paper.

\subsection{Trotterized evolution within digital-analog approach}

The feature of superconducting quantum processor architecture, described in the preceding subsection, can be exploited to simulate physical systems in a non-standard way.
The idea is to use the internal physics of the quantum chip to model a Trotterized evolution of quantum systems, described by Hamiltonians with interactions digitally reducible to ZZ interaction.
Many spin Hamiltonians contain a part of nearest-neighbor spin ZZ interaction (or XX interactions or combination of XX, YY, and ZZ terms), so the evolution generated by this part of Hamiltonian can be transferred to physics of the quantum processor.
Using this approach, it is possible to simulate the dynamics of spin models (such as the Ising model and the  Heisenberg model) with arbitrary single spin terms of the Hamiltonian. A change of the basis is needed to simulate XX, YY, or ZX interaction, which is possible due to the individual control of qubits (single-qubit gates).

To illustrate our idea, we focus on the most straightforward example -- simulation of spin clusters, described by transverse-field Ising model with Hamiltonian
\begin{equation}
    H =  -\sum_{j}h_{j}\sigma^{x}_{j} -\sum_{<ij>}J_{ij}\sigma^{z}_{i}\sigma^{z}_{j},
    \label{trIsing}
\end{equation}
where $h_{j}$ are spin local fields and $J_{ij}$ are interaction constants. This model requires to use only single-qubit gates to simulate the effect of terms $h_{j}\sigma^{x}_{j}$ since the ZZ part of the Hamiltonian is realized by coupling between qubits of the quantum processor. If we identify every qubit with spin, the influence of crosstalks on qubits states will be similar to the evolution of spins under the ZZ interaction hamiltonian.

We notice that the topology of the quantum processor dictates the topology of the spin system we can simulate, using this approach (at least, in a straightforward version we here implement). Ratios between values of $J_{ij}^{phys}$ of the processor fix ratios between $J_{ij}$ of the simulated spin system. Nevertheless, parameters $h_{j}$ of the simulated system can be arbitrary. Moreover, by applying additional Pauli-X gates, it is possible to switch off interaction of certain qubits of the device and thus also to engineer digitally the topology of the spin cluster.

In order to trace the free evolution of the system, the Trotter decomposition can be used to split an evolution operator for a full Hamiltonian, consisting of non-commuting operators, into a sequence of evolution operators for individual contributions.
The simplest form of this decomposition for the Hamiltonian $H = H_{A} + H_{B}$ is
\begin{equation}
      e^{-it(H_{A}+H_{B})} \approx (e^{-iH_{A}\frac{t}{N_{tr}}}e^{-iH_{B}\frac{t}{N_{tr}}})^{N_{tr}},
\end{equation}
which in $N_{tr} \rightarrow \infty$ limit becomes exact.  For transverse field Ising Hamiltonian, an appropriate splitting of the full Hamiltonian is
\begin{equation}
    H_{A} = -\sum_{j}h_{j}\sigma^{x}_{j},
\end{equation}
\begin{equation}
    H_{B} = -\sum_{<ij>}J_{ij}\sigma^{z}_{i}\sigma^{z}_{j},
\end{equation}
where the latter term corresponds to ZZ interaction between qubits. Notations $h_{j}$, $J_{ij}$, and $t$ refer to the system we wish to simulate (i.e., not to the actual parameters of the quantum processor). The ratios between $J_{ij}$ and constants of ZZ interactions $J^{phys}_{ij}$ of qubits of the chip must be the same, although the absolute values for the two sets of these quantities can be, of course, different.

In terms of quantum schemes, every two-qubit ZZ-rotation operator within the Trotterized evolution can be performed without application of any quantum gate from the universal set and just idling for some time instead, as depicted schematically in  Fig. 1. The upper scheme in Fig. 1(a) shows usual representation of the two-qubit operator
$$
\exp\left(-ih_i t \sigma_{i}^{x}\right) \exp\left(-ih_j t \sigma_{i}^{x}\right) \exp\left(-i J_{ij}t \sigma_{i}^{z} \otimes \sigma_{j}^{z}\right)
$$
using the quantum circuit language (fully digital approach). For more details on the the structure of this quantum circuit see Refs. \cite{weare,Simulation3}. The lower scheme in Fig. 1(a) shows the same operator within our digital-analog approach. The operator $\exp\left(-i J_{ij}t \sigma_{i}^{z} \otimes \sigma_{j}^{z}\right)$ is implemented in the analog way by idling within some physical time interval. The physical idling time $t^{phys}$ is given by the relation $J_{ij}t = J^{phys}_{ij}t^{phys}$, so the controlling dimensionless parameters are $J^{phys}_{ij}t^{phys}$ sometimes referred to as Ising times. To implement such a time delay in IBM Q machines, one needs to incorporate a particular number of identity gates $I$ into the circuit. This number denoted as $M$ can be calculated from a simple relation $M = t^{phys}/T_{I}$, where $T_{I}$ is a physical time of the identity gate (in IBM Q machines, $T_{I} \approx 100$  ns and it also can vary from chip to chip). The technique can be applied to each Trotter step for the evolution operator and the whole structure of the circuit both for digital and digital-analog approaches is shown schematically in Fig. 1(b).

In our experiments, we simulate several spin systems and focus on mean of excitations as a function of time. The operator of excitation number for a single spin $j$ is $n_j = \sigma^{+}_{j}\sigma^{-}_{j}$. The operator of averaged over spins excitation number reads $n = \frac{1}{N_q}\sum_{j}n_j$. The quantity we are interested in is the expectation value $<n>\equiv \bra{\psi} n \ket{\psi}$, which we refer to as the mean excitation number.

The main steps of our digital-analog simulations are:
\begin{enumerate}
    \item Prepare an initial state of the quantum register of $N_{q}$ qubits, for example, $\ket{0}^{\otimes N_{q}}$;
    \item Apply Trotter-decomposed evolution operator to a prepared register of qubits; the construction of the evolution operator in terms of IBM Q gates is different for digital and digital-analog approaches, as explained above;
    \item Measure the state of the qubit register in computational basis and thus obtain a sample of $N_{runs}$ bitstrings of the form, e.g., "01001".
    \item Using the sample, calculate the mean excitation number of the spin system as
    \begin{equation}
        <n(t^{phys})> = \frac{1}{N_{runs}}\sum_{k=1}^{N_{runs}}\sum_{j=1}^{N_{q}}s^{j}_{k}(t^{phys})
    \end{equation}
    where $s^{j}_{k}(t^{phys})$ is a $j$-th qubit measurement outcome, obtained in the $k$-th run of the quantum circuit with a time parameter $t^{phys}$.
\end{enumerate}
\begin{figure}[h!]
\includegraphics[width=1.0\linewidth]{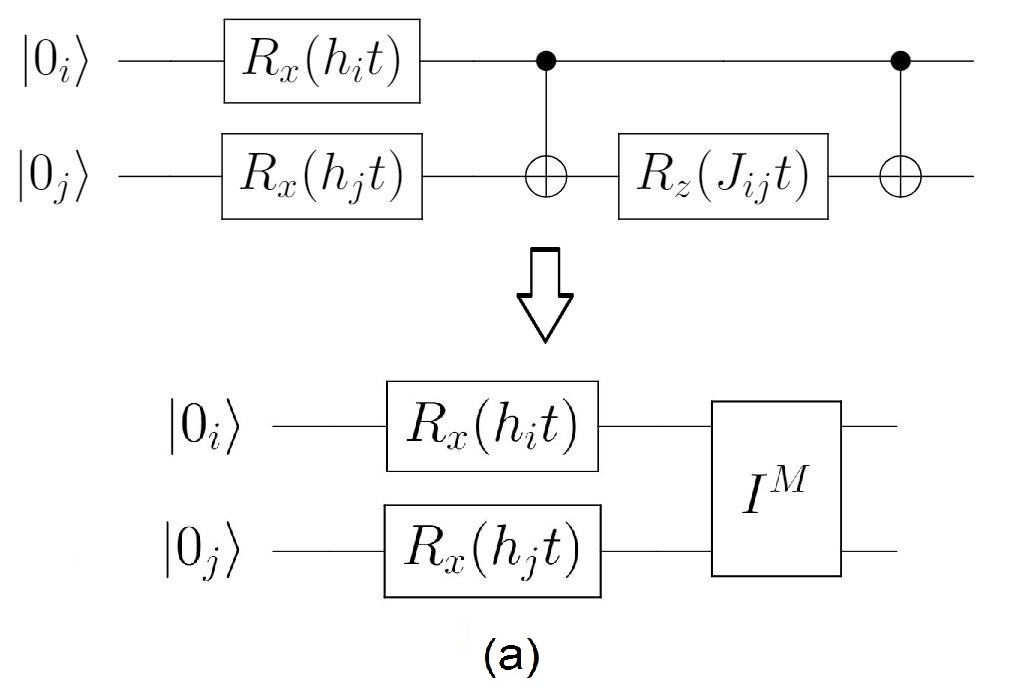}
\includegraphics[width=1.0\linewidth]{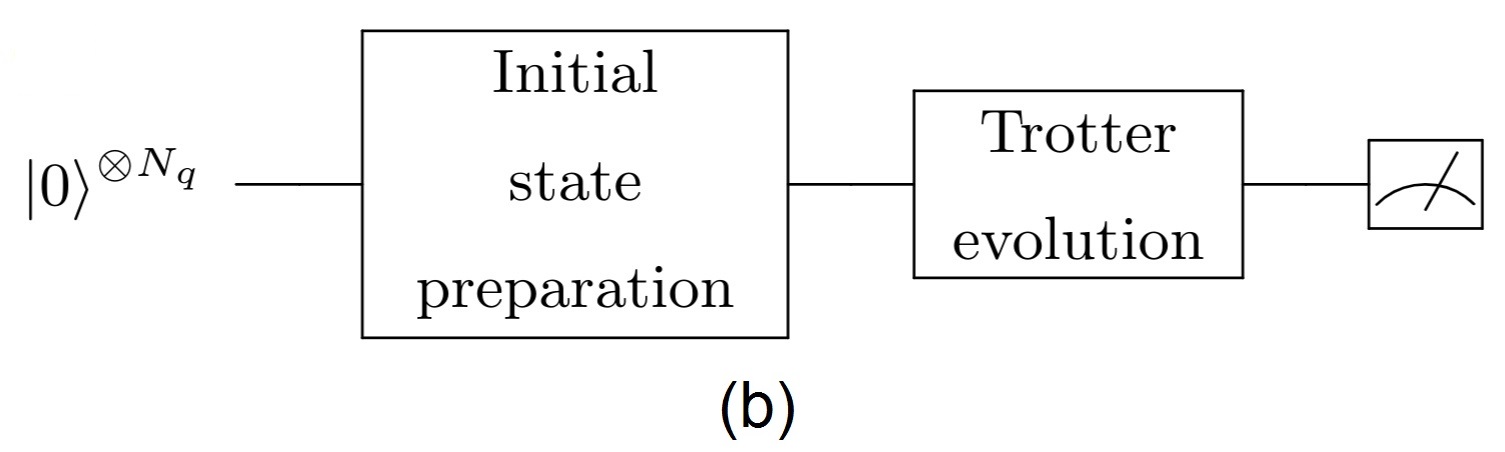}
\caption{(a) A single Trotter step for a couple of spins of the transverse field Ising model. The upper picture represents a standard digital decomposition into quantum gates. The lower picture represents the same subroutine based on the ZZ interaction of the qubits of the quantum chip. (b) A general schematic of quantum circuit to simulate spin systems dynamics.}
\end{figure}




\section{Main results}

To illustrate our ideas, we perform several experiments for different physical setups. The first experiment deals with the dynamics of a transverse field Ising model for just two spins as a toy model. Every spin is represented by a qubit of the quantum processor, and the Trotterized evolution is realized using both digital and digital-analog approaches. Then, we increase the number of spins in the system (up to 5 and 14) to see how the growth of the circuit influences results, and compare the accuracy of computation for these two approaches. Finally, we digitally introduce artificial disorder into the 14-spin cluster and simulate the dynamics.

\subsection{Toy model: two spins}

For the case of two qubits of a quantum processor, our approach can more accurately reproduce dynamics of transverse field Ising system for initial basis states (e.g. $\ket{00}$ and $\ket{11})$ compared to the standard digital approach based on CNOT gates. The particular topology for the part of the IBM QX14 processor, used in the experiment, is shown in Fig. 2, where we utilized qubits $Q_{0}$ and $Q_{1}$ for this simulation. Numbers near edges between qubits show corresponding values of $J^{phys}_{ij}$ measured in kilohertz\cite{ibm_device}.

We used 6-step Trotter decomposition to study the dynamics of the two-qubit system. The number of Trotter steps was chosen through a comparison with the exact solution of the Schrodinger equation; for six steps, the Trotterization error is generally not significant for the evolution time we consider, which is limited by $T_{1,2}$ (except the long evolution times close to the upper bound). Notice that we do not pay attention to this error since we aim to analyze errors, associated with the hardware imperfections. Thus, we change $t^{phys}$, and, for each value of $t^{phys}$, we divide it into six intervals and implement six Trotter steps. We have chosen this strategy to be able to characterize the ability of quantum devices to run quantum algorithms within two distinct approaches (digital and digital-analog methods), i.e., we focused on the characterization of the device imperfections. In other words, we are mostly interested in infidelities between perfect and imperfect realizations of the same circuit. Another approach is to analyze and infidelity between the full time-dependent solution of the Schr\"{o}dinger equation and the Trotterized realization with the real quantum device. We believe that such an analysis can be considered as a complementary investigation for the path we follow. We also would like to note that the issue of a proper Trotterization is rather complicated even if we disregard the errors of a quantum device. This is particularly true for disordered systems, see, e.g., Ref. \cite{Heyl}. Quantum models we simulate are also disordered due to different values of couplings between qubits. This is another reason why we would like to focus on errors of realizations with real devices.

The maximum $t^{phys}$ in our simulation of the two-spin model was around 50 microseconds, which nearly corresponds to both $T_{1}$ and $T_{2}$ for a chosen quantum processor -- apparently, this is the main restriction of our approach. Maximum dimensionless Ising time $J_{01}t=J_{01}^{phys}t^{phys}$ was equal to approximately 2.5. In our simulations, we assumed that every spin is subjected to the homogeneous transverse field. Both parameters $h_{0}$ and $h_{1}$ are equal to $2J_{01}$ that corresponds to the regime when ZZ interaction and transverse field are of nearly equal importance.

In all experiments, we estimated statistical error for the mean excitation number using the standard error for a sample of experimental data $n_i$:
\begin{equation}
    SE = \sqrt{\frac{1}{N_{runs}(N_{runs}-1)}\sum_{i=1}^{N_{runs}}(n_{i} - \overline{n})^{2}},
\end{equation}
where $\overline{n}=\frac{1}{N_{runs}}\sum_{i=1}^{N_{runs}}n_{i}$, which for $N_{runs} = 8192$ gives SE = $\mathcal{O}$ $(10^{-3})$.

\begin{figure}[h!]

    \center{\includegraphics[width=0.5\linewidth]{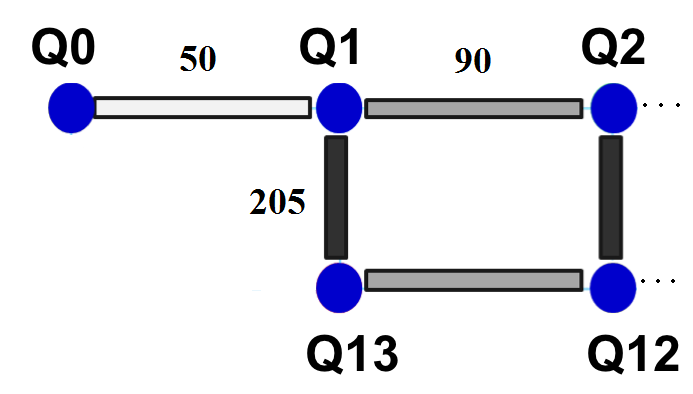}}
    \caption{A schematic view of a part of the IBM QX14 quantum processor, utilized for the simulation of the two-spin system ($Q_{0}$ and $Q_{1}$ are used). CNOT gates can be applied directly between qubits, which are connected by lines. Those qubits experience pairwise ZZ stray interaction of Ising type due to the presence of resonator they share. Numbers between qubits show corresponding values of coupling constants measured in kilohertz.}

\end{figure}

Both experimental and theoretical results for the mean number of excited spins $<n(t)>$ are presented in Fig. 3. It shows a good agreement between the theory and the results of the digital-analog simulation. The theoretical result is based on an approximation of the same level (6-step Trotter decomposition). For the purely digital evolution, the agreement between the theory and the experiment is much poorer. To characterize the quality of experiments, we calculated several figures of merit (see Appendix A for more detail), which support our general conclusions.

Note that the theoretical curve demonstrates an irregular behavior in Fig. 3 at long times, where it starts to deviate significantly from the results of the digital-analog simulation. This feature is related to the fact that at such long times Trotterization error becomes significant and a larger number of Trotter steps is needed. In general, Trotterization errors are manifested through irregular and unstable dynamical behavior. In contrast, decoherence smears out such features within the digital-analog strategy. Also, note that the initial time in our digital simulations was not exactly zero, but we used a certain small value (0.001 in terms of the mean Ising time). This was done on purpose since otherwise the IBM compiler optimizes the circuit automatically at zero time by removing quantum gates from the circuit, which leads to the unphysical jump of $<n(t)>$ at $t = 0$.

We emphasize that all actual connections of every chosen qubit to other qubits of the quantum processor should be taken into account in our simulations because even if neighboring qubits ($Q_{2}$ and $Q_{13}$ in this simulation) stay in ground states, they are still connected to qubits used in the experiment ($Q_{1}$) and influence their time evolution.
Thus, we perform our simulation for the four-spin cluster ($Q_{0}$, $Q_{1}$, $Q_{2}$, and $Q_{13}$), where two spins play a passive role.
We compare experimental data with the theory, taking into account the presence of surrounding qubits.

\begin{figure}[h!]

    \center{\includegraphics[width=1.0\linewidth]{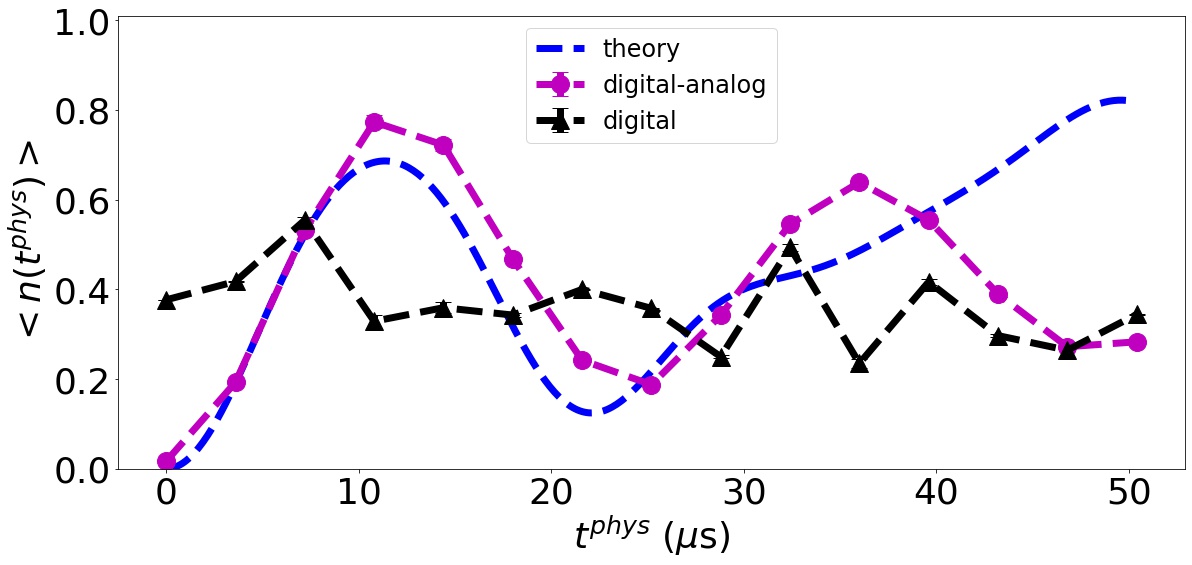}}
    \caption{Dynamics of the mean excitation number for two-spin transverse field Ising cluster during the free evolution from the initial state $\ket{00}$. Shown are results of the same approximation level (6 Trotter steps) for theoretical calculation (blue color), for the experiment within the digital-analog approach (magenta color), and the experiment within the digital approach (black color). Physical time $t^{phys}$ defines the total time of algorithm execution within the digital-analog approach, which gives rise to ZZ interaction of qubits. For both theoretical results and the result of experimental implementation of the digital approach, $t^{phys}$ must be mapped on time $t = J^{phys}_{ij}t^{phys}/J_{ij}$ of the simulated system.}

\end{figure}

\begin{figure}[h!]
\includegraphics[width=0.5\linewidth]{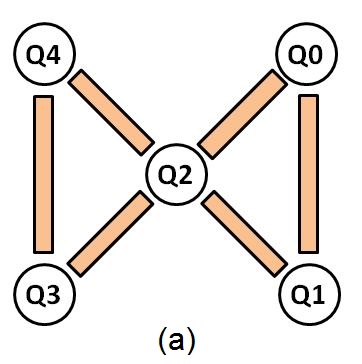}
\includegraphics[width=1.0\linewidth]{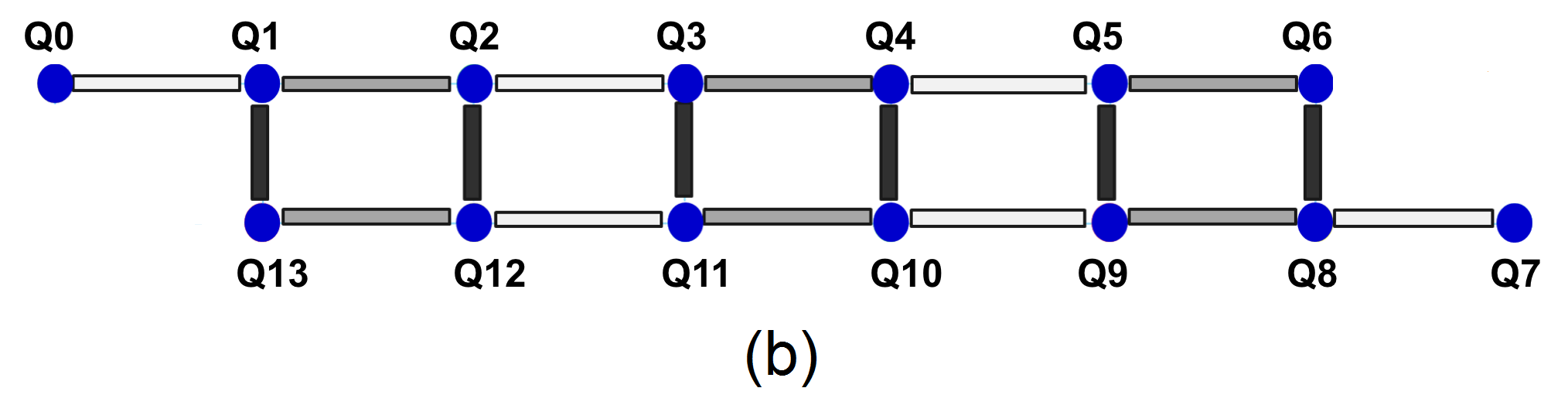}
\caption{A topology of IBM QX2 (a) and IBM QX14 (b) quantum processors, which determine the connectivity map of spin clusters we simulate.}
\end{figure}




\subsection{Dynamics of 5-spin and 14-spin clusters}

To test our idea further, we used IBM QX2 and IBM QX14 processors. Full connectivity maps of these processors are provided in Fig. 4. We exploited all qubits of these quantum chips to model a temporal evolution of spin clusters.
We again assumed homogeneous field in Hamiltonian (1) -- all parameters $h_{j}$ are equal to the averaged over the cluster $2J_{ij}$.

In these experiments, a Trotter decomposition of evolution operator for the transverse field Ising model requires many two-qubit gates within the digital strategy, leading to significant error accumulation, see, e.g., Refs. \cite{weare,Simulation3}. Figure 5 shows both experimental and theoretical results for the dynamics of the mean number of excited spins in 5- (a) and 14-spin (b) clusters. The number of Trotter steps was 6 and 3, respectively. Thus, this figure compares results of three types, but with the same Trotter numbers: (i) using a unitary simulator of the quantum processor (theory), (ii) using a digital-analog approach, and (iii) using a digital approach.

From this figure, we can see that a conventional digital approach completely fails to reproduce theoretically expected dynamics with the chosen number of Trotter steps. The reason is that every native two-qubit gate has an error of several percents, and to model a dynamics of 5 and 14 spin clusters, one needs to use many two-qubit gates. At the same time, if we replace two-qubit gates with the evolution of the system under crosstalks' influence, then there is a good correspondence between theoretical and experimental results. Notice that for the 14-spin experiment, we chose Trotter number 3 because, within the digital approach, it already gives results, which are even qualitatively incorrect.

We remind that the important timescale in our simulation is given by the average Ising times $\overline{J_{ij}^{phys}} \overline{T_1}$ and $\overline{J_{ij}^{phys}} \overline{T_2}$. For the IBM QX2 processor, these times are slightly larger, which makes this quantum machine more suitable for the hybrid approach. It is seen from Fig. 5 (a) that the time dependence of mean excitation number for 5-spin cluster includes also a second oscillation and not only the first one, in contrast to similar experimental data for the 14-spin cluster, see Fig. 5 (b). This is due to a larger mean crosstalk value in the 5-qubit machine. We see that experimental results obtained within the digital-analog approach both for 5- and 14-spin clusters reproduce systematically faster evolution compared to the theoretical results. A possible explanation is that some crosstalk parameters used in our simulations \cite{ibmqx1,ibm_device} are underestimated compared to the actual ones.

In Ref. \cite{Simulation1} authors used quantum hardware with embedded interaction between qubits and investigated a phenomenon of many-body localization in an XY spin model without the usage of Trotterization. Single-qubit gates have been used only to create an initial highly excited state, while the interaction between qubits produced a nontrivial dynamics. The Trotterized evolution has an advantage which originates from the flexibility of digitization -- the effect of various additional (tunable) terms of the modeled Hamiltonian, not present in the actual interaction of qubits in the chip, can be simulated, as well as time-dependent Hamiltonians. The disadvantage of this strategy is associated with additional digitization errors.

It is interesting to explore deeper an impact of flexibility, as provided by digitization, for the implementation of other quantum algorithms, which are different from quantum simulation of spin clusters with interaction terms of the Hamiltonian similar to the interaction of qubits of the device. To this end, in Appendix B, we consider a more "mathematical" problem, which is quantum Fourier transform with three qubits of IBM QX2 processor. For this particular quantum circuit, the results of digital-analog and digital implementation with available IBM Q machines have nearly the same accuracy. This implies that specialized quantum processors designed for digital-analog computation schemes potentially can outperform digital quantum computers for such problems as well.  In Section IV we discuss an issue of optimal parameters for such processors in a more detail.





\begin{figure}[h!]
\includegraphics[width=1.0\linewidth]{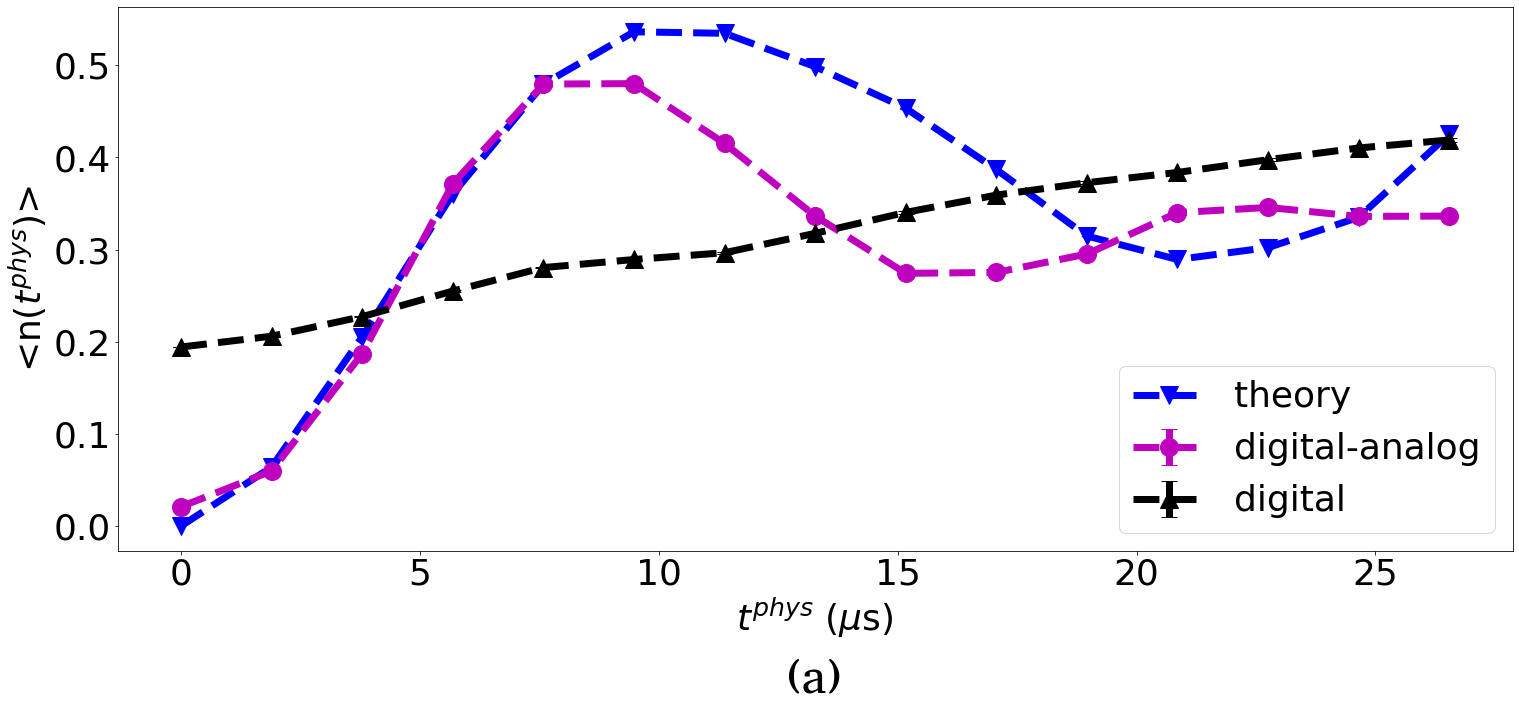}
\includegraphics[width=1.0\linewidth]{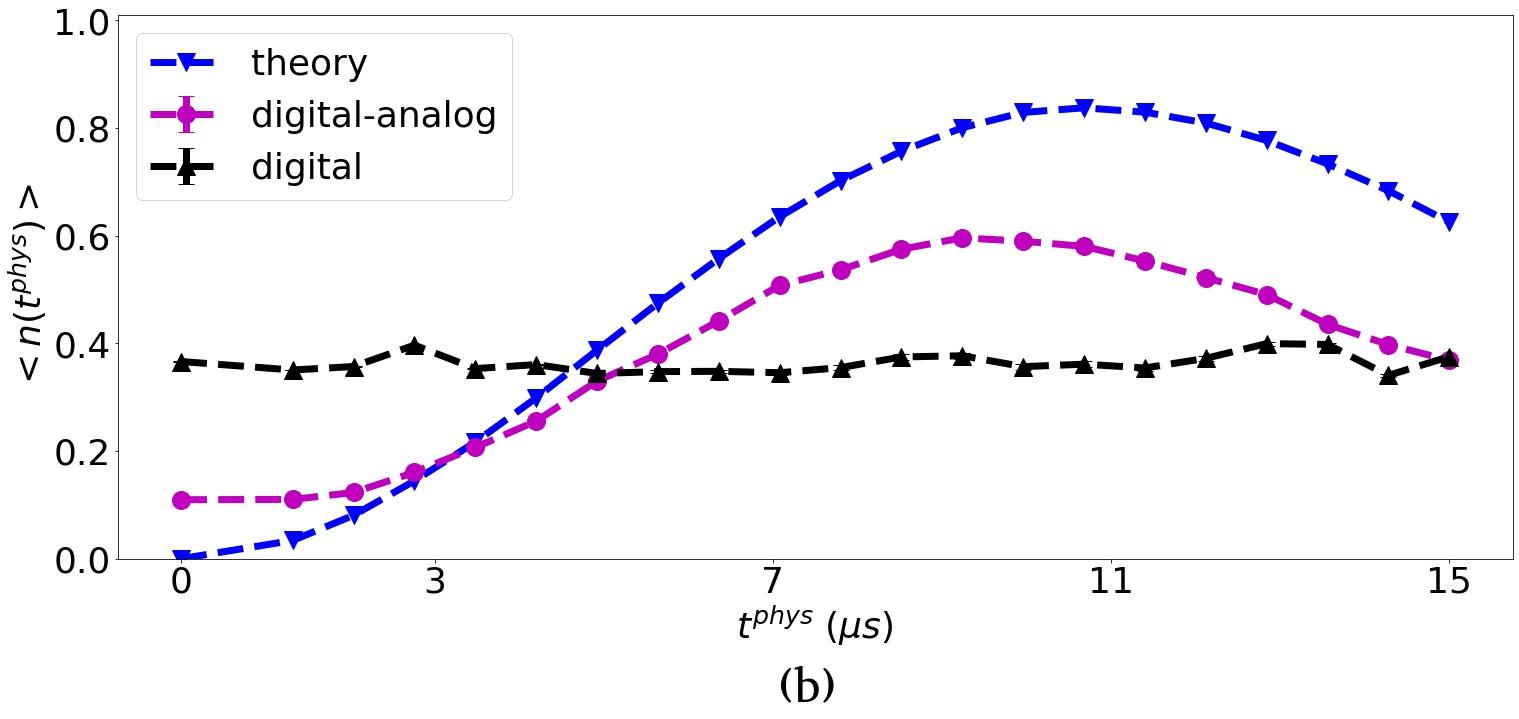}
\caption{Dynamics of the mean excitation number for 5-spin and 14-spin transverse field Ising systems during the free evolution from the initial state $\ket{00..0}$. Shown are the results of the same approximation level (6 Trotter steps for 5-spin system and 3 Trotter steps for the 14-spin system) obtained theoretically (blue color), experimentally within the digital-analog approach (magenta color) and experimentally within the digital approach (black color). The experiments for 5- and 14-spin clusters were performed with IBM QX2 and IBM QX14 processors, respectively. Physical time $t^{phys}$ defines the total time of algorithm execution within the digital-analog approach. For both theoretical and experimental implementation results of the digital approach, $t^{phys}$ must be mapped on time $t = J^{phys}_{ij}t^{phys}/J_{ij}$ from the simulated model.}
\end{figure}

\subsection{Simulation of the effect of disorder}

In this subsection, we incorporate digitally additional disorder to the Hamiltonian (\ref{trIsing}), which refer to the random magnetic field in $z$ direction, $\epsilon_j \sigma_j^{z}$. Thus, two types of single-spin terms are introduced to the Hamiltonian, which corresponds to two different directions of the magnetic field. This model is known to have a rich phase diagram \cite{Imbrie} showing a tendency towards localization in the large spin number limit as amplitudes of the magnetic field in $z$ direction increases or alternatively, as its amplitude in $x$ direction decreases. The single-body terms of the above form are simulated by adding corresponding single-qubit rotations around $z$ axis within each Trotter step. For this experiment, we chose an IBM QX14 processor to make the modeled systems large enough to incorporate many-body phenomena. This processor has crosstalk values in a wide range, thus already having a disorder in parameters. We assume that artificial disorder parameters $\epsilon_j$ are randomly chosen from a uniform distribution in the interval [$-2\overline{J}$,  $2\overline{J}$], $\overline{J}$ being an average interaction between all spins of the system.

 The initial state of the 14-spin system is shown schematically in Fig. 6. It contains a long domain wall between spin-up and spin-down regions, which is expected to decay faster in absence of an artificial additional disorder. In our experiments, we focus on the time evolution of the initial magnetization pattern and the half-difference of spin magnetization in the chosen pattern. The operator of the former quantity is defined as $m_j=2n_{j}-1$, while the operator of the latter quantity is
\begin{equation}
    n_{half} = \sum_{j \in S^{init}_{up}}\frac{1}{N^{init}_{up}}m_j - \sum_{j \in S^{init}_{down}}\frac{1}{N^{init}_{down}}m_j,
\end{equation}
where $S^{init}_{up}$ and $S^{init}_{down}$ are sets of spins with initial magnetization +1 ("up") and -1 ("down"), while $N^{init}_{down}$ and $N^{init}_{up}$ are their numbers, correspondingly. Figure 7 shows the experimental results for the expectation values of these operators.
Figure 7 (a) depicts typical patterns of magnetization as a function of time in the absence and the presence of the disorder (a particular realization of the disorder is chosen for the right panel). Figure 7 (b) shows the time dependence of the half-difference of spin magnetization, averaged over 10 disorder realizations. We see that the presence of the additional disorder extends the survival time of the initial domain wall, as expected. However, we cannot enter the most interesting regime of large Ising times, since the maximum evolution time is limited by $T_{1,2}$, which exceeds typical entangling time $1/\overline{J_{ij}^{phys}}$ only by a factor of 3-4, so our results can be considered as only qualitative and illustrative. Notice that the fact that two curves in Fig. 7 (b) start approaching each other at $t^{phys}\gtrsim 40$ $\mu s$ is due to the digitization errors, which become noticeable at corresponding values of Ising time.

\begin{figure}[h!]

    \center{\includegraphics[width=1.0\linewidth]{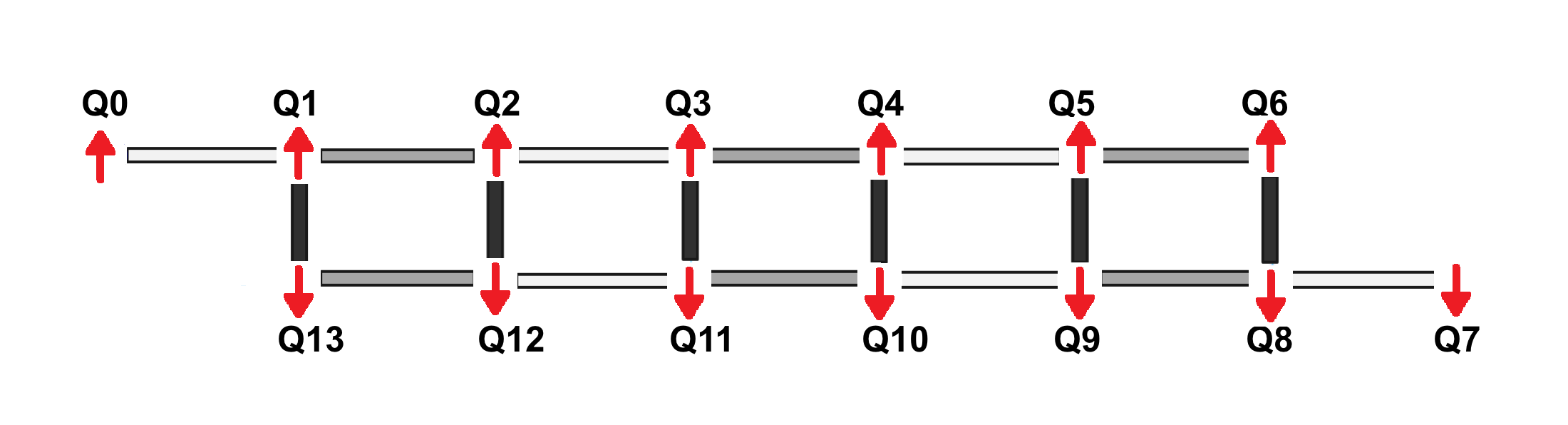}}
    \caption{Initial states of the spin cluster. The free evolution of the system from this state is studied.}

\end{figure}

\begin{figure}[h!]
\includegraphics[width=1.0\linewidth]{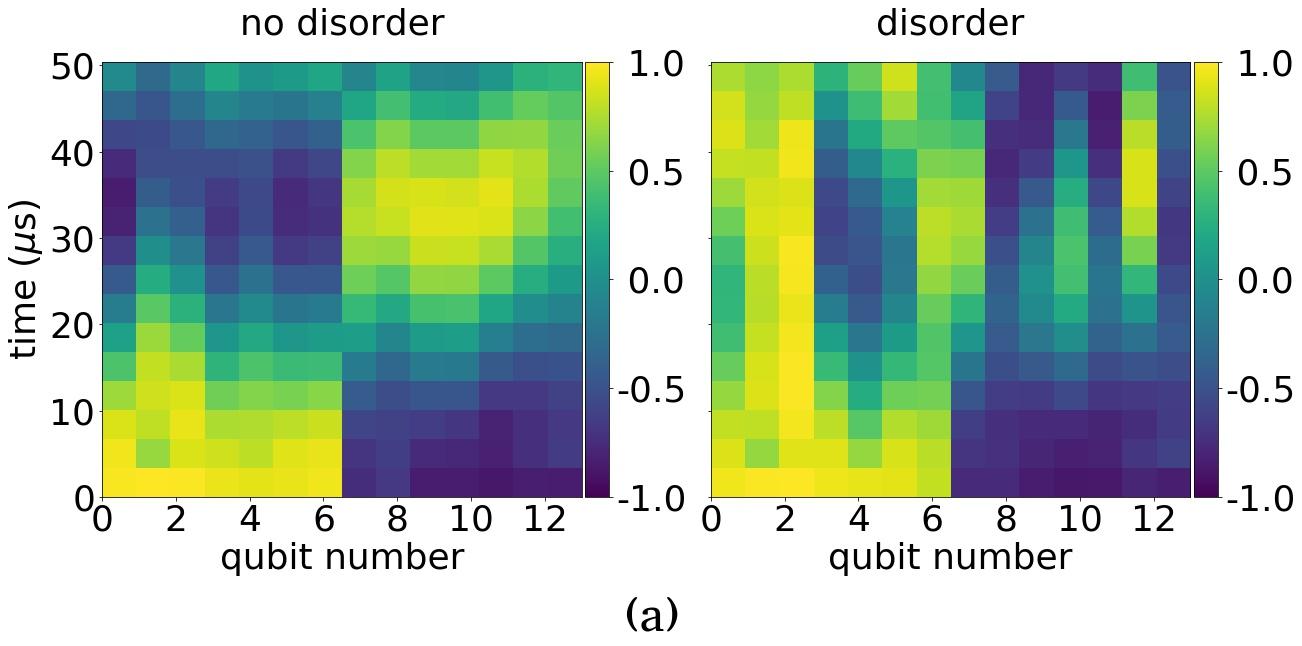}
\includegraphics[width=1.0\linewidth]{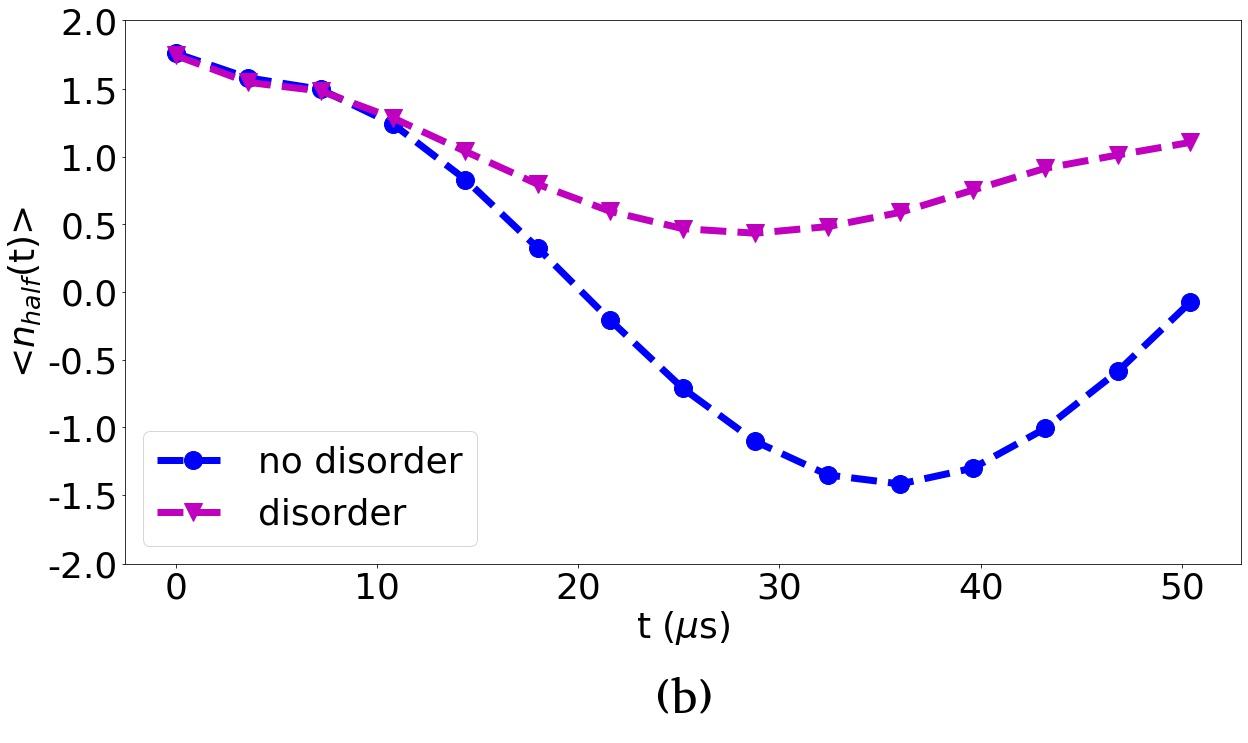}
\caption{(a) A pattern of magnetization, calculated in the absence of disorder (left panel) and the presence of disorder (right panel) on IBM QX14 quantum processor over the time for a 14-spin cluster during the free evolution under the transverse field Ising Hamiltonian (6 Trotter steps). The colorbar corresponds to spin magnetization value (b) Half-difference spin magnetization values over time in the presence (blue) and the absence (magenta) of the disorder.}
\end{figure}






\section{Summary and discussion}

In this paper, we argued that it is prospective to use a digital-analog quantum computation strategy within near-term NISQ technology. There can be various modifications of this approach, but the general idea is to avoid the use of two-qubit gates of the universal set and to rely on physical interaction between qubits, which can naturally bring entanglement into the quantum system.

The most straightforward application of this approach is quantum simulation of spin systems. Particularly, always-on interaction between qubits of the quantum processor can be used to realize interaction between spins within Trotterized (discretized) evolution, while single-qubit gates can be applied to introduce any desired single-qubit terms of the Hamiltonian as well as to possibly modify interaction terms by changing basis. More technically advanced approach is to use tunable couplers to switch off interaction during execution of single-qubit gates.

We illustrate our ideas with the quantum processors of IBM Quantum Experience, which are constructed for digital (not for digital-analog) computation. Nevertheless, qubits of these machines experience residual static couplings known as crosstalks which physically originate from interaction between qubits through virtual microwave photons. These couplings are responsible for additional (correlated) errors within digital quantum computing paradigm, which are harder to correct. They produce non-Markovian dynamics of qubits and one of the examples of such a dynamics is analyzed both experimentally and theoretically in the Appendix C. Although negative effect of crosstalks is known from literature, see, e.g., Ref. \cite{crosstalk}, our simple illustration shows explicitly that the actual computation time in NISQ devices may be limited not by $T_{1,2}$, but by crosstalks ($\sim 1/J^{phys}$).

In contrast to digital approach, in our digital-analog simulations, we rely on this always-on interaction due to crosstalks. Despite of the fact that crosstalks are small (tens or hundreds of kilohertz), the results of the digital-analog simulation of spin clusters dynamics in the transverse-field Ising model are much more accurate than similar results for the conventional digital simulation based on two-qubit CNOT gates. The latter approach suffers from severe error accumulation problem and yields a very low accuracy under just few Trotter steps, while the digital-analog approach still gives results accurate on semi-quantitative level.

There are other computational problems, one can approach with the digital-analog strategy and in principle it can be used to realize any unitary transformation \cite{SB,DAQC2}. In Appendix B we applied this strategy for the quantum Fourier transform with three qubits available IBM quantum machines and compared the result with the results of digital quantum computation. For this particular problem, accuracies of both implementations turn out to be nearly the same. We attribute this result mainly to the fact that crosstalks in IBM Q machines are too small to efficiently realize controlled rotations on large angles, which are necessary in quantum Fourier transform, since decoherence starts to play significant role on such time scales (see Appendix B for a more detail).

We would like to stress that our experimental results have been obtained using quantum machines designed to keep crosstalks between qubits small, since they produce additional errors in digital quantum computation. Let us now estimate a value of coupling constant between qubits, which is optimal for digital-analog computation based on always-on interaction, which alters fidelity of single-qubit gates. We assume that coupling energy of a pair of qubits is $J$, so that the typical entangling time for a pair of qubits is $t_{2q} \sim 1/J$. Typical duration $t_{1q}$ of single-qubit gates for IBM Q machines is nearly 50 ns. The error of the single-qubit gate due to the coupling between the qubits can be estimated as $\sim Jt_{1q}$. Due to the decoherence, the entangling operation is also not ideal and it is characterized by the error $\sim t_{2q}/T_{1,2} \sim 1/JT_{1,2}$, where $T_{1,2}$ is typically 50-100 $\mu s$. Within this model, the total error of the single step of the algorithm (single-qubit gate plus idling) is of the order of $Jt_{1q}+1/JT_{1,2}$. The optimal value of $J$ which corresponds to the minimum of this expression is $J_{opt}\sim 1/\sqrt{t_{1q}T_{1,2}}$, while the minimum of the total error is $\sim \sqrt{t_{1q}/T_{1,2}}$. For the parameters of IBM Quantum Experience, we obtain an estimate $J_{opt}\sim 1$ MHz which is one or two orders of magnitude larger than the typical crosstalk value in IBM Q machines and it can be achived using both the capacitive coupling and the coupling through a resonator.

Thus, in our illustrations with IBM Q devices, we are far away from the optimal regime for digital-analog quantum computation strategy. Yet our experimental results are much more accurate than the results of standard digital approach for a very specialized problem, which is a simulation of transverse-field Ising model, or have nearly the same accuracy for a less specialized and more "mathematical" problem, which is quantum Fourier transform with three qubits. This shows that it is prospective to construct specialized quantum processors designed for digital-analog simulation, particular quantum algorithms and possibly particular topology. Being in optimal regime, one can implement $\sim J_{opt}T_{1,2}$ entangling steps, which is nearly 50-100 for the above values of $t_{1q}$ and $T_{1,2}$. We remind that the actual crosstalk values in our experiments were small, so that the number of such steps was an order of magnitude smaller. Of course, there are also other methods to increase the efficiency of digital-analog computation, which can be seen in the increase of $T_{1,2}$ and decrease of $t_{1q}$ thus making an error $\sqrt{t_{1q}/T_{1,2}}$ smaller. In principle, it is also possible to extend the time of simulation beyond $T_{1,2}$ by using a dynamical decoupling within each idling time interval. However, this requires a modified estimate for $J_{opt}$, since the number of single-qubit gates per Trotter step would be enhanced; this issue is beyond the scope of the present paper. Another potentially useful technique is related to the zero-noise extrapolation \cite{mitig1}. For aforementioned specialized processors it could be used to mitigate errors of single-qubit gates.

Our results also suggest that for the digital-analog computation with superconducting qubits coupled by resonators, it is preferable not to switch to the new "rotated" basis, as done in IBM Q setup. This transform suppresses the crosstalk, which starts to be determined residually by third levels of transmons (not the direct physical interaction involving two lowest levels), and increases the fidelity of single-qubit gates. The above discussion shows that it is favorable to operate in the regime at which errors of the single-qubit gates caused by the always-on coupling and errors of entangling operations caused by decoherence are of the same order. Note that Without an additional rotation, coupling between the qubits would be of XX and not of ZZ form.

Of course, our estimates for the optimal couplings are valid only for digital-analog strategy based on always-on interaction. For architectures based on adjustable couplers, allowing for the isolation of qubits during single-qubit gates, our simple model is not applicable, so that stronger couplings might be preferable leading to faster entangling operations. In this context, we would like to mention a recent paper \cite{foxen2020demonstrating}, which reports on realization of a continuous set of two-qubit excitation-conserving gates appropriate for fermionic simulation problems \cite{Babbush}. The use of these gates instead of the standard gates from the universal set is able to enhance significantly compilation efficiency, while conservation of excitation number allows to eliminate a need for microwave control. Certain analogies can be noticed with the approach we follow in our experiments, since we also get rid of microwave control for a continuous set of entangling ZZ operations, which conserve excitation number and have immediate application in quantum simulation of spin models. Apparently, the approach of Ref. \cite{foxen2020demonstrating} is promising for the construction of specialized quantum processors appropriate for quantum simulation of both spin models and fermionic systems in NISQ era. It is of interest to explore a potential of digital-analog strategy based on always-on interaction for the simulation of fermionic systems along the ideas of Ref. \cite{Babbush}.

Let us mention that the newer generation of IBM Q devices is characterized by CNOT gates infidelities, which are significantly lower than infidelities of the devices used by us, and therefore accuracy of digital simulation are expected to be higher compared to the illustrative results shown in this paper.

Note that the hybrid digital-analog approach can be also exploited in variational quantum-classical calculations. Indeed, always-on interaction can be used as an entangler and durations of "waiting" intervals can be considered as additional variational parameters. Similar but fixed entangler was used in Ref. \cite{variat} to evaluate molecular energies. This entangler was constructed from a fixed sequence of microwave pulses rather than from always-on interaction between qubits. Of course, other physical realizations of quantum computers or simulators different from superconducting quantum circuits, can be also prospective for the implementation of digital-analog strategy.



\section*{Acknowledgements}
We acknowledge use of the IBM Quantum Experience for this work. The viewpoints expressed are those of the authors and do not reflect the official policy or position of IBM or the IBM Quantum Experience team.

D. V. B. and A. A. Zh. acknowledge a support from RFBR (project no. 20-37-70028). W. V. P. acknowledges a support from RFBR (project no. 19-02-00421).



\FloatBarrier

\renewcommand{\appendixname}{Appendix}
\appendix

\section{Figures of merit}

To quantify results from Sec. III.A, we use several figures of merit.
We analyze $l_{1}$ metric defined as
\begin{equation}
    l_{1} = |<n^{exp}> - <n^{theory}>|,
\end{equation}
where $<n^{exp}>$ and $<n^{theory}>$ are experimental and theoretical values of the mean excitation number, correspondingly. The resulting plot is provided in Fig. 8, which shows better accuracy of the digital-analog approach compared to the standard digital approach. As the physical simulation time approaches $T_1$ and $T_2$, the agreement between the theory and the results of our approach also becomes unsatisfactory.

We also perform a frequency analysis of $<n(t)>$. The Fourier transform of $<n(t)>$ is presented in Fig. 9., the Fourier components of $<n(t)>$ are
\begin{equation}
    n(\omega_{k}) = \sum_{m = 0}^{n-1}\tilde{n}(t_{m})\exp\biggl(-2\pi i \frac{t_{m}\omega_{k}}{n}\biggl), \textbf{    } k = 0, ..., n-1,
\end{equation}
where $\tilde{n}(t_{m})$ are normalized values of $<n(t)>$, and
\begin{equation}
    \tilde{n}(t_{m}) = \frac{n(t_{m})}{\max\limits_{t_{m}} n(t_{m})},
\end{equation}
where $t_{m}$ is a moment of evolution time of the excitation measurement.

Figure 9 illustrates that noisy two-qubit gates increase the zero-frequency component, i.e., effectively flatten time dependence of $<n(t)>$ besides changing its time variation character. We have already seen such a tendency towards flattering of various quantities evaluated using noisy quantum hardware as functions of controlling parameters \cite{QML_paper}.

\begin{figure}[h!]

    \center{\includegraphics[width=1.0\linewidth]{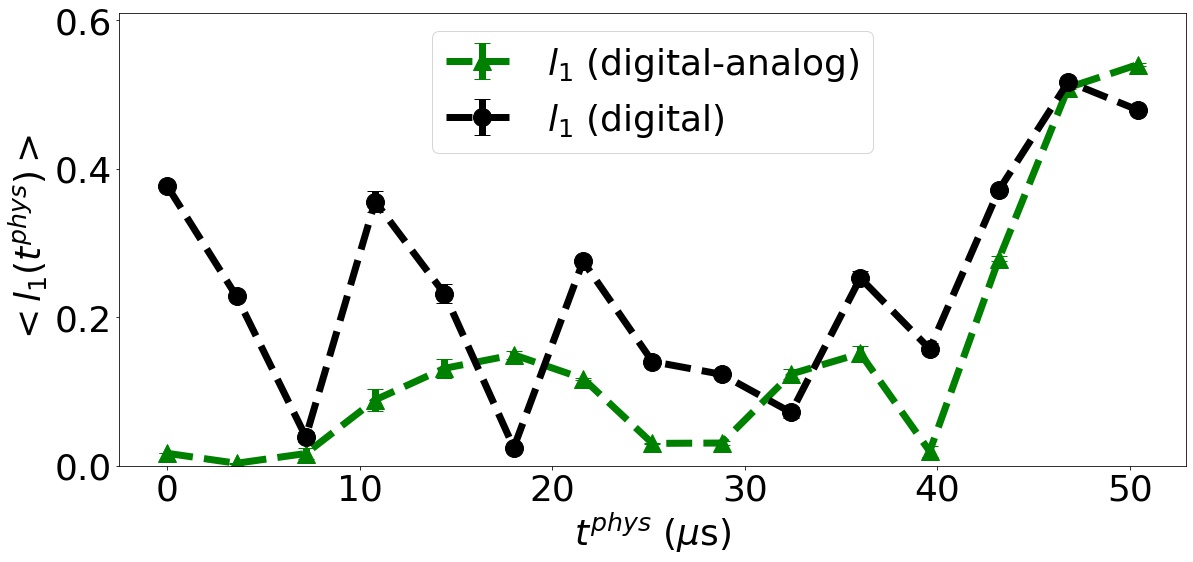}}
    \caption{Resulting plots of averaged over two qubits $l_{1}$ metric for two-qubit initial state $\ket{00}$ as a function of time during the free evolution. The experimental results within both the digital-analog approach and within the digital approach are shown.}

\end{figure}

\begin{figure}[h!]

    \center{\includegraphics[width=1.0\linewidth]{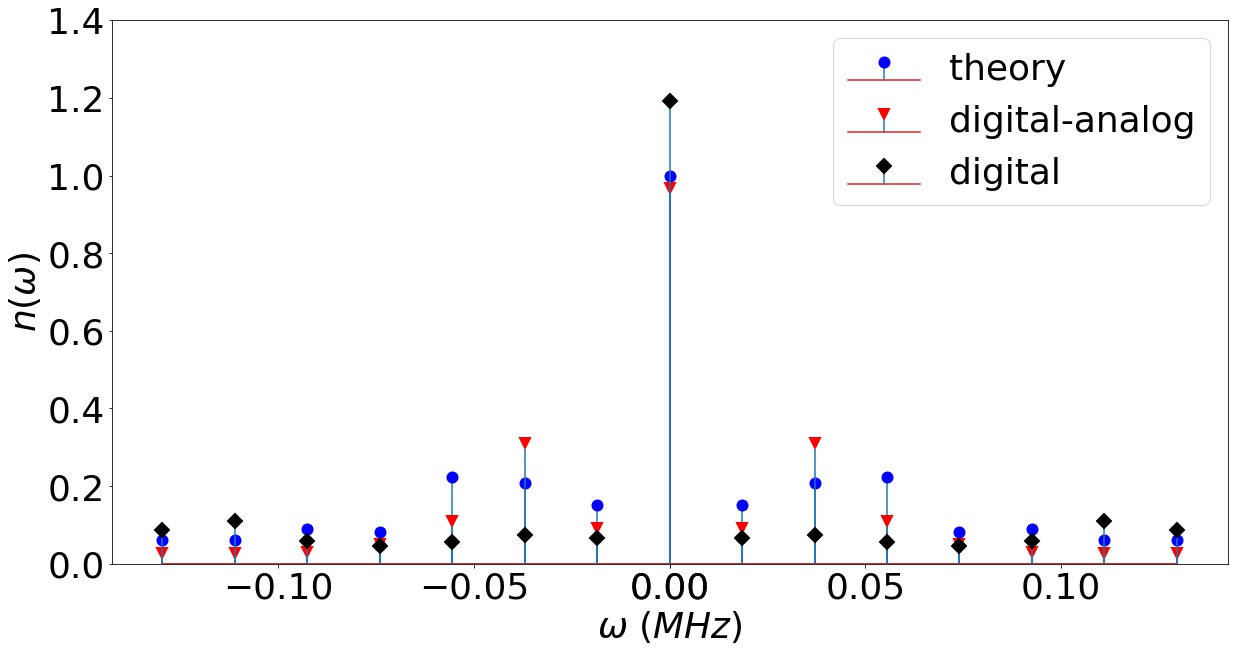}}
    \caption{Fourier components of the mean excitation number as a function of time.}

\end{figure}


\section{Quantum Fourier transform}

To provide an example of a "more mathematical"  problem, which can be approached with the digital-analog approach, we consider here the quantum Fourier transform. It is an important quantum computing subroutine, providing super-polynomial speed-up concerning classical analog and is a part of other quantum algorithms (e.g., Shor prime factorization, quantum phase estimation).

The quantum Fourier transform is defined as
\begin{equation}
    \mathcal{F}\ket{\Omega} = \frac{1}{\sqrt{N}}\sum_{k=0}^{N-1}e^{2\pi i\Omega k/N}\ket{k},
\end{equation}
where $N$ is the number of basis states. The operation $\mathcal{F}$ can be decomposed in the composition of Hadamard gates and controlled phase rotation gates, where the latter operation is
\begin{eqnarray}
    cR_{k} = \ket{0}\bra{0}\otimes\mathcal{I} + \ket{1}\bra{1}\otimes R_{k} =
    \notag \\
    =
    \begin{pmatrix}
    1 & 0 & 0 & 0 \\
    0 & 1 & 0 & 0 \\
    0 & 0 & 1 & 0 \\
    0 & 0 & 0 & e^{2\pi i/2^{k}}.
    \end{pmatrix}
\end{eqnarray}
 This gate can be constructed from single and two qubit operations, which are parts of the Ising model Hamiltonian. For a couple of qubits 1 and 2, described by the Hamiltonian
\begin{equation}
    H = J_{12}\sigma^{z}_{1}\sigma^{z}_{2},
\end{equation}
the controlled phase rotation gate has the following expression in terms of spin operations
\begin{equation}
    cR_{k} = e^{-iJ_{12}\sigma^{z}_{1}\sigma^{z}_{2} t}e^{iJ_{12} \sigma^{z}_{1} t}e^{iJ_{12} \sigma^{z}_{2} t},
\end{equation}
where $t=\dfrac{\pi}{2^{k}J_{12}}$ is system evolution time. Hereafter in this Appendix we omit label "phys" in all formulas.

To implement the quantum Fourier transform, using the digital-analog approach, on a quantum processor with constant qubit-qubit interactions, one needs to take into account a simultaneous phase accumulation of different qubits and to use proper phase shifts to compensate undesired accumulated phases. This can be accomplished with standard single-qubit gates. In the following, we demonstrate an example of quantum Fourier transform calculation for a simple system of three qubits.

Consider a system of three qubits of a quantum processor with constant qubit-qubit interactions, described by the Hamiltonian
\begin{equation}
    H_{int} = J_{12} \sigma^{Z}_{1}\sigma^{Z}_{2} + J_{13} \sigma^{Z}_{1}\sigma^{Z}_{3} + J_{23} \sigma^{Z}_{2}\sigma^{Z}_{3}.
\end{equation}
The evolution operator is
\begin{eqnarray}
    e^{-i(J_{12}\sigma^{Z}_{1}\sigma^{Z}_{2} + J_{13}\sigma^{Z}_{1}\sigma^{Z}_{3} + J_{23}\sigma^{Z}_{2}\sigma^{Z}_{3})t} =
    \notag \\
    = e^{-iJ_{12}\sigma^{Z}_{1}\sigma^{Z}_{2}t}e^{-iJ_{13}\sigma^{Z}_{1}\sigma^{Z}_{3}t}e^{-iJ_{23}\sigma^{Z}_{2}\sigma^{Z}_{3}t}
\end{eqnarray}
To implement a controlled phase rotation of one qubit under the control of the another one, we need to get rid of the third qubit influence with the effect of spin echo: we apply a $\pi$-pulse (the NOT gate) at half of evolution time to turn back the evolution of the third qubit state

\begin{eqnarray}
    \notag \\
    \sigma^{X}_{3}e^{-iH_{int}t}\sigma^{X}_{3} =
    e^{-iJ_{12}\sigma^{Z}_{1}\sigma^{Z}_{2}t}
    e^{iJ_{13}\sigma^{Z}_{1}\sigma^{Z}_{3}t} \times
    \notag\\
    \times e^{iJ_{23}\sigma^{Z}_{2}\sigma^{Z}_{3}t}.
\end{eqnarray}

\noindent The full time evolution operator with the spin echo is
\begin{equation}
    e^{-iH_{int}t}\sigma^{X}_{3}e^{-iH_{int}t}\sigma^{X}_{3} = e^{-iJ_{12}\sigma^{Z}_{1}\sigma^{Z}_{2}t}.
\end{equation}
If we add single-qubit phase rotations, similar to (B4), we will implement a controlled phase rotation between two chosen qubits. Thus, one can realize every part of the quantum Fourier transform with the digital-analog approach. The circuit, implementing the controlled phase rotation gate with the digital-analog approach, is shown in Fig. 10.

\begin{figure}[h!]
    \includegraphics[width=1.0\linewidth]{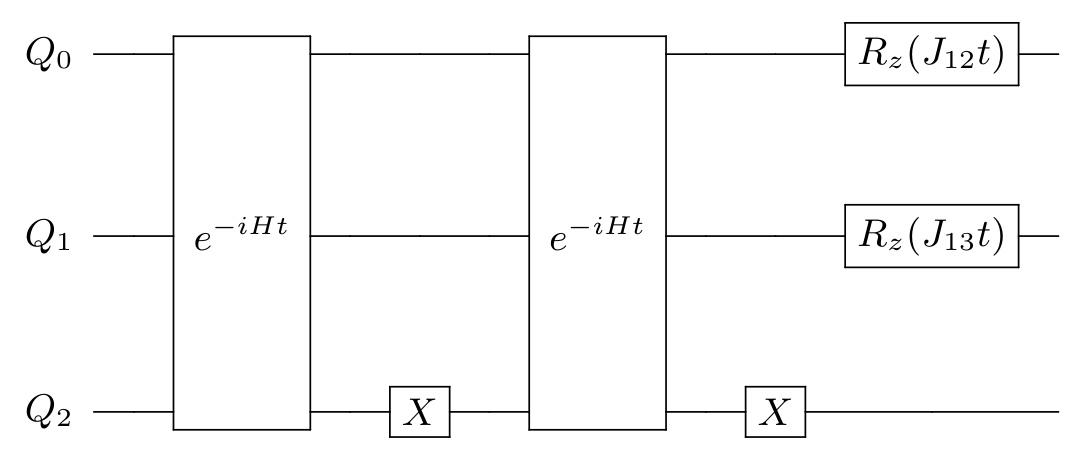}
    \caption{A schematic of the controlled phase rotation gate, implemented with the digital-analog approach. Two qubit gates are realized with simultaneous evolution of three qubits and echo operation, applied to qubit $Q_{2}$ to turn back the evolution of the third qubit state.}
\end{figure}

To characterize performance of two approaches, in the following we use two measures - a classical trace distance and a Bhattacharya distance. Let X is a random variable, which takes values from a set $\mathcal{X} = \{x\}$. For two probability distributions $\{p_{x}: x \in \mathcal{X}\}$ and $\{q_{x}: x \in \mathcal{X}\}$, the trace distance is
\begin{equation}
    D(\{p_{x}\}, \{q_{x}\}) = \frac{1}{2}\sum_{x \in \mathcal{X}}|p_{x} - q_{x}|
\end{equation}
and Bhattacharyya distance is
\begin{equation}
    F(\{p_{x}\}, \{q_{x}\}) = -ln\biggl(\sum_{x \in \mathcal{X}}\sqrt{p_{x}q_{x}}\biggl)
\end{equation}.

To compare the digital-analog approach with the digital, we run experiments with all initial states of the computational basis. The outcomes of these experiments are probabilities of measuring qubits in basis states, $\{p^{ideal}_{i}: i = 000, 001, ..., 111\}$ for the experiment on a simulator and $\{p^{exp}_{i}: i = 000, 001, ..., 111\}$ for the experiment on a quantum processor. In Fig. 11, we show trace distance and Bhattacharya distance values for experiments with quantum Fourier transform, applied to different initial states.
 According to the first metric, digital approach is slightly more accurate, while according to the second metric digital-analog method is slightly better. We conclude that the accuracy of the digital-analog approach is early the same as the accuracy of digital implementation of the quantum Fourier transform. Having used only hard-wired qubit-qubit interactions and having no possibility to fine-tune them to optimize the performance of the algorithm, we obtained good results, which encourage further investigation of opportunities to use the digital-analog quantum computation scheme (see Section IV of the main text for the discussion).

\begin{figure}[h!]
\includegraphics[width=1.0\linewidth]{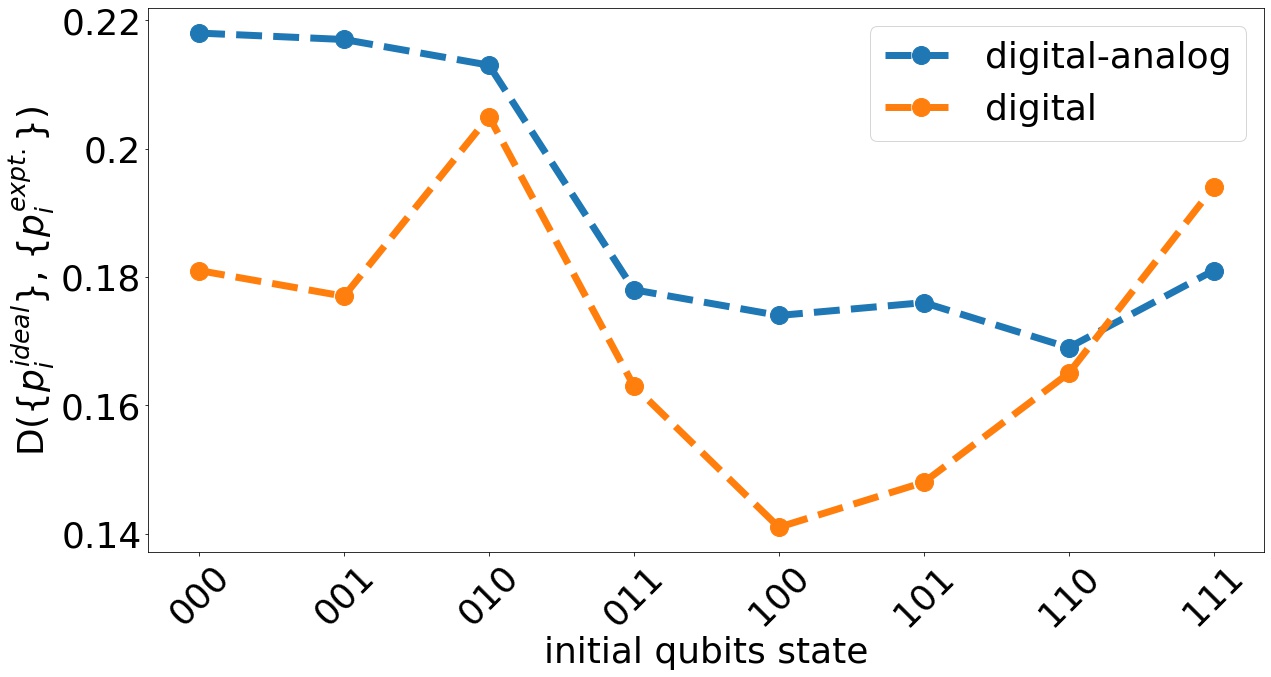}
\includegraphics[width=1.0\linewidth]{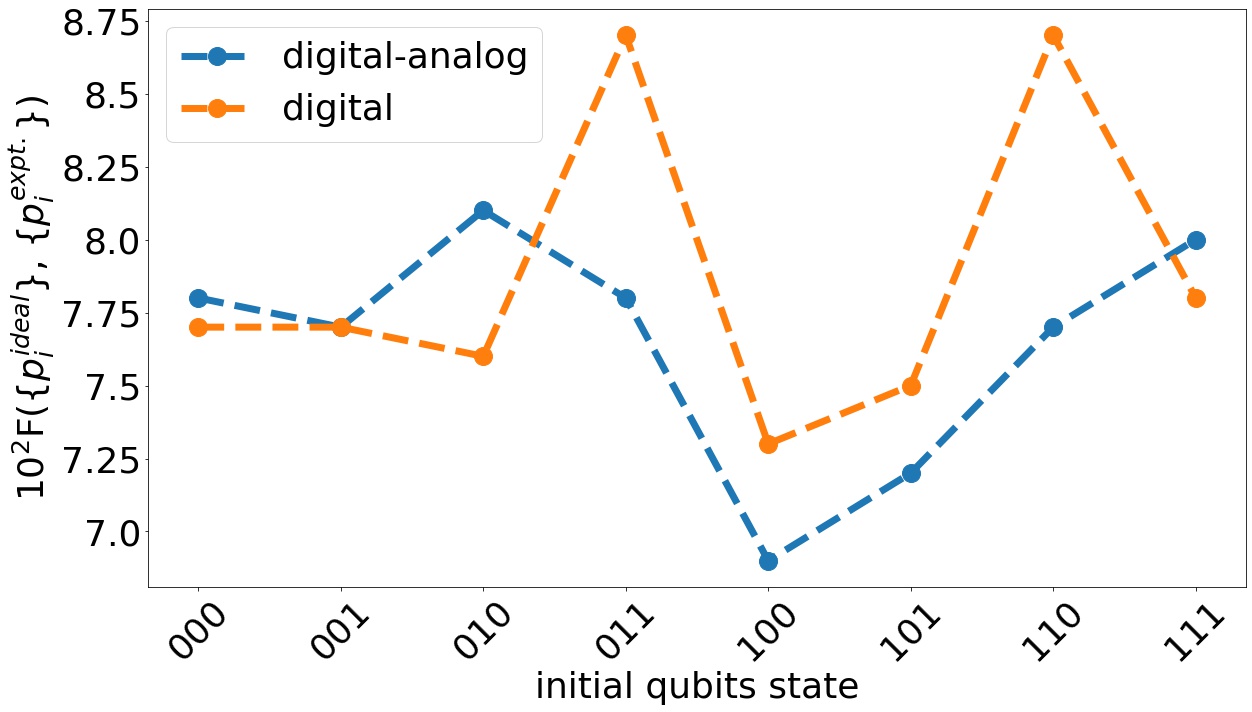}
\caption{Trace distance (a) and Bhattacharyya distance (b) values for different initial states after quantum Fourier transform.}
\end{figure}

Thus, we see that the digital-analog implementation of quantum Fourier transform does not outperform digital implementation so significantly as in the case of quantum simulation of transverse-field Ising model. We believe that there are two major reasons for this. Firstly, although on a circuit level the quantum Fourier algorithm is simple, nevertheless, it includes few controlled rotations on large angles that means that idling time within digital-analog approach is long. This makes the result sensitive to decoherence processes. Secondly, in contrast to the quantum simulation problem, this quantum circuit contains only few CNOTs and therefore typical infidelity of digital realization is not very high. Let us stress that digital-analog approach turns out to be more efficient than the digital realization provided (i) the quantum circuit contains multiple CNOTs, (ii) the topology of the quantum chip is suitable for the particular problem within the digital-analog approach. This shows that it is prospective to fabricate specialized processors for the digital-analog strategy.


\section{Non-Markovian dynamics of qubits}

In the main text, we used constant qubit-qubit interaction between superconducting qubits to simulate the dynamics of different spin systems. In the digital quantum computation with these physical systems, stray coupling leads to non-Markovian dynamics, which introduces unwanted phase shift to qubit states and thus serves as an additional source of computational error.
In this appendix, we provide an example of non-Markovian dynamics in IBM QX4 quantum processor and discuss how it affects computation. In particular, we compare the characteristic time of non-Markovian system state revival with average coherency time of processor qubits.

The investigation of non-Markovian dynamics of quantum systems is an active field of research in open quantum systems physics. This kind of system behavior plays a role in many problems, such as spin squeezing \cite{Yim2012}, the transition of quantum systems to classical state \cite{Xiong2015}, the assistance of quantum information processing algorithms \cite{Dong2018}, and an application of quantum physics to photosynthesis \cite{Chen2015}.

While working with non-Markovian systems, one of the main issues is characterizing the degree of non-Markovianity.
In complex systems, there are usually many environments, and depending on the environment state and the nature of the interaction, the dynamics of the system can be Markovian or non-Markovian.
The most straightforward way to characterize an open system is to write down the master equation for the system-bath hamiltonian and then introduce simplifications from physical reasons \cite{Breuer2007}.
This approach lacks universal simplicity, because it requires to derive the master equation for every single system, and does not provide a unified measure of non-Markovianity degree.

To quantify the behavior of open quantum systems, different measures of non-Markovianity were introduced (see Refs. \cite{Nonmarkovianity1}, \cite{Chan2016}, \cite{He2017}). In the following, we use a trace distance between two initial states of the system under the scope, following ideas from Ref. \cite{Nonmarkovianity1}.
The trace distance between two quantum states $\rho_{1}$ and $\rho_{2}$ defines as following:
\begin{equation}
    D(\rho_{1}, \rho_{2}) = \frac{1}{2}Tr\sqrt{(\rho_{1} - \rho_{2})^{\dagger}(\rho_{1} - \rho_{2})}.
\end{equation}
The operational meaning of this measure follows from a formula
\begin{equation}
    Pr(\rho_{1}, \rho_{2}) = \frac{1 + D(\rho_{1}, \rho_{2})}{2},
\end{equation}
which is a probability to successfully distinguish between two states $\rho_{1}$ and $\rho_{2}$, given one of them as an output from a black-box source.

\begin{figure}[h!]

    \center{\includegraphics[width=0.5\linewidth]{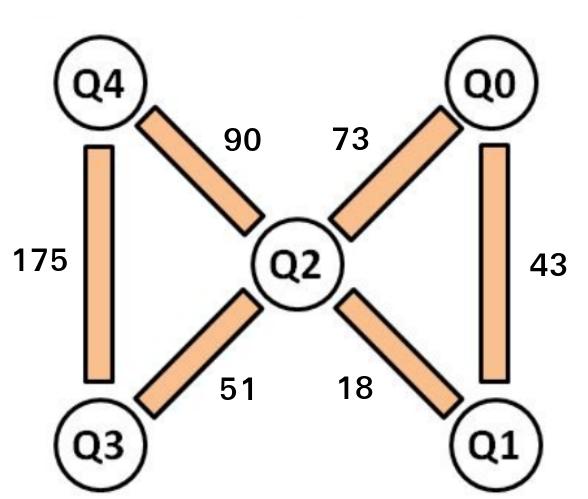}}
    \caption{A scheme of IBM QX4 quantum processor. CNOT gates can be applied directly between those qubits, which are connected by orange lines. Qubits, connected by these lines, experience pairwise ZZ interaction of Ising type due to the presence of resonator they share. Numbers, placed between qubits, show corresponding values of coupling constants measured in kilohertz.}

\end{figure}

In our experiment, we considered the evolution of a superconducting transmon qubit of the IBM QX4 processor in the presence of crosstalk interaction with surrounding qubits. We used one of the boundary qubits of the five-qubit processor as the target system (e.g., qubit $Q_{0}$ in Fig. 12), so there are two neighboring qubits. The Hamiltonian of the target qubit interaction with neighbors is
\begin{equation}
    H_{zz} = J_{01}^{phys}\sigma^{z}_{0}\sigma^{z}_{1} + J_{02}^{phys}\sigma^{z}_{0}\sigma^{z}_{2} + J_{12}^{phys}\sigma^{z}_{1}\sigma^{z}_{2},
\end{equation}
where $J^{phys}_{ij}$ are qubit-qubit crosstalks. Let us derive the expression for a trace distance between two initial states of the target qubit. The initial state of the target qubit with environment qubits is
\begin{equation}
    \ket{\Psi(0)} = \ket{S}\ket{E},
\end{equation}
where the environment state has the form
\begin{equation}
    \ket{E} = \alpha\ket{00} + \beta\ket{01} + \gamma\ket{10} + \delta\ket{11}.
\end{equation}
During the evolution, the entanglement between the target qubit state with the neighboring qubits state leads to loss of the target state coherency, bringing it closer to a maximally mixed state. If two states of the target qubit were initially distinguishable, they then lose their distinguishability due to interaction with the environment.
The character of this loss defines the information dynamics in the system. If the distinguishability drops monotonically, then the evolution is Markovian, otherwise, it is non-Markovian.
Because the trace distance characterizes how distinct two quantum states are, one can use it to estimate the open quantum system dynamics (see Refs. \cite{Nonmarkovianity1}, \cite{Nonmarkovianity}, \cite{Nonmarkovianity2}). To reveal those effects, one should use states, which are maximally distinct at the initial moment of the experiment \cite{Nonmarkovianity3}.

We consider the dynamics of the target qubit for initial states from the computational and the Hadamard bases. The evolution of these states during time $t^{phys}$ under $H_{zz}$ leads to the transformation
\begin{equation}
    \rho_{0}(0) =
    \begin{pmatrix} 1 & 0 \\  0 & 0  \end{pmatrix}
    \longrightarrow
    \rho_{0}(t^{phys}) =
    \begin{pmatrix} 1 & 0 \\  0 & 0  \end{pmatrix},
\end{equation}
\begin{equation}
    \rho_{1}(0) =
    \begin{pmatrix} 0 & 0 \\  0 & 1  \end{pmatrix}
    \longrightarrow
    \rho_{1}(t^{phys}) =
    \begin{pmatrix} 0 & 0 \\  0 & 1  \end{pmatrix},
\end{equation}
\begin{eqnarray}
    \rho_{\pm}(0) =
    \begin{pmatrix} \frac{1}{2} & \pm\frac{1}{2} \\  \pm\frac{1}{2} & \frac{1}{2}  \end{pmatrix}
    \longrightarrow
    \rho_{\pm}(t^{phys}) =
    \notag \\
    = \begin{pmatrix} \frac{1}{2} & \pm B \\  \pm B^{*} & \frac{1}{2}  \end{pmatrix},
\end{eqnarray}
where
\begin{eqnarray}
    B = e^{-i\omega_{0}t^{phys}}
    \biggl(
    \frac{|\alpha|^{2}}{2}e^{-2i(\tau_{01}+\tau_{02})} +
    \notag \\
    \frac{|\beta|^{2}}{2}e^{-2i(\tau_{01}-\tau_{02})} +
    \frac{|\gamma|^{2}}{2}e^{2i(\tau_{01}-\tau_{02})} +
    \notag \\
    \frac{|\delta|^{2}}{2}e^{2i(\tau_{01}+\tau_{02})}
    \biggl),
\end{eqnarray}
with notation $\tau_{ij} = \omega_{ij}t^{phys}$ introduced for the simplicity.
We see that for the initial quantum state with a trivial relative phase there is no influence from surrounding qubits through ZZ-interaction. On the other side, when the initial state has a non-trivial quantum phase, there is a strong influence on the target qubit dynamics. For the latter case, using the formula (C1) and expressions of density matrix after the evolution (C8), we obtain the expression for trace distance
\begin{equation}
    D = 2\sqrt{BB^{*}}
\end{equation}
with
\begin{eqnarray*}
BB^{*} =
\frac{|\alpha|^{2}+|\beta|^{2}+|\gamma|^{2}+|\delta|^{2}}{4} +
\notag \\
\frac{|\alpha|^{2}|\beta|^{2} + |\gamma|^{2}|\delta|^{2}}{2}\cos\tau_{02} +
\notag \\
\frac{|\alpha|^{2}|\gamma|^{2} + |\beta|^{2}|\delta|^{2}}{2}\cos\tau_{01} +
\notag \\
\frac{|\alpha|^{2}|\delta|^{2}}{2}\cos(\tau_{01}+\tau_{02}) +
\notag \\
\frac{|\delta|^{2}|\gamma|^{2}}{2}\cos(\tau_{01}-\tau_{02}).
\end{eqnarray*}

Let us choose four states from Bell basis as initial state of the neighboring qubits. The Bell states are
\begin{eqnarray}
\ket{\Phi_{\pm}} = \frac{1}{\sqrt{2}}(\ket{00} \pm \ket{11}),\notag \\
\ket{\Psi_{\pm}} = \frac{1}{\sqrt{2}}(\ket{10} \pm \ket{01}).
\label{Bell}
\end{eqnarray}
For there states, using the formula (C8) ($\alpha$ = $\delta$ = $\frac{1}{\sqrt{2}}$, $\beta$ = $\gamma$ = 0 for $\ket{\Phi_{\pm}}$ or $\alpha$ = $\delta$ = 0, $\beta$ = $\gamma$ = $\frac{1}{\sqrt{2}}$ for $\ket{\Psi_{\pm}}$) we obtain the expression for the trace distance:
\begin{equation}
    D(t^{phys}) = |\cos[2(J^{phys}_{01} \pm J^{phys}_{02})t^{phys}]|.
\end{equation}

In our experiments with the IBM QX4, we used qubit $Q_{0}$ as a target qubit and $Q_{1}$ and $Q_{2}$ qubits as neighbor qubits. For the initial states pair ($\ket{+}$, $\ket{-}$) of the target qubit and the initial state of the neighboring qubits, $\ket{\Phi_{+}}$, dynamics of the trace distance over the time is illustrated in Fig.13.

The theoretical curve is plotted for the time dependence of trace distance without taking into account finite single-qubit coherency times $T_{1}$ and $T_{2}$. We see that, after about five microseconds, two initial states of the target qubit become indistinguishable. This time is an order less than an average coherency time of IBM QX4 processor (about 50$\mu s$), which means that the influence of the crosstalk on the qubit state evolution is stronger than single-qubit decoherence.

We observed similar results for other initial Bell states of the neighboring qubits (although for $\ket{\Psi_{+}}$, $\ket{\Psi_{-}}$ states the oscillation period is about twice larger) for the target qubit initial states ($\ket{+}$, $\ket{-}$).
At the same time, for initial states ($\ket{0}$, $\ket{1}$), there was no such oscillation dependence.

From these simple observations, we estimate the influence of crosstalk interaction between superconducting fixed-frequency qubit as a limiting factor for quantum computation within the considered architecture. As most problems require operating with quantum states with a non-trivial quantum phase, one should take this interaction into account in the form of additional error correction.

\begin{figure}[h!]

    \center{\includegraphics[width=1.0\linewidth]{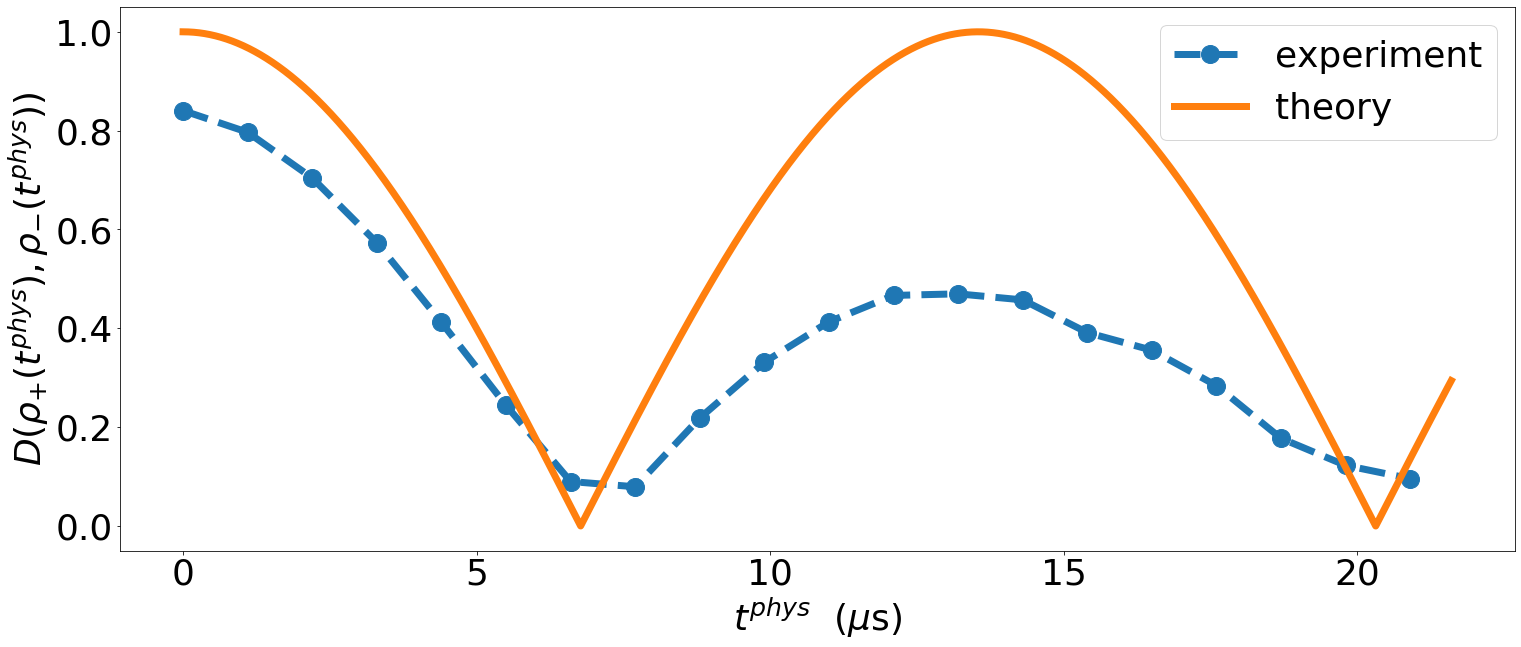}}
    \caption{A plot of trace distance over time for a $Q_{0}$ target qubit and  $Q_{1}$ and $Q_{2}$ neighboring qubits. Trace distance is calculated for qubit $Q_{0}$ initial states $\ket{+}$ and $\ket{-}$. The surrounding qubits $Q_{1}$ and $Q_{2}$ were prepared in a maximally entangled state $\ket{\Phi_{+}}$. The orange curve corresponds to the exact calculation of trace distance dynamics, and the blue curve corresponds to calculation on a quantum processor.}

\end{figure}


\bibliographystyle{apsrev4-2}
\bibliography{ref}

\end{document}